\documentclass[a4paper,fleqn,useAMS,usenatbib]{mnras}

\pdfoutput=1

\usepackage{mathptmx}

\usepackage[T1]{fontenc}
\usepackage{ae,aecompl}

\usepackage[dvips]{graphicx}
\usepackage{amsmath}
\usepackage{amsfonts}
\usepackage{longtable}
\usepackage{amssymb}
\usepackage{comment}
\usepackage[dvipsnames]{xcolor}
\usepackage[%
        ]{hyperref}

\hypersetup{
	colorlinks=true,	
	urlcolor=MidnightBlue,		
	pdfpagelabels=true,
	hypertexnames=true,
	plainpages=false,
	naturalnames=true,
	pdftitle={New Periodic TESS Signals},    
	pdfauthor={Chakraborty et al.},     
	linkcolor=WildStrawberry,          
	citecolor=ForestGreen,        
}

\renewcommand{\textbf}{}

\newcommand{\Kepler}{{\it Kepler}}
\newcommand{\TESS}{{\it TESS}}

\newcommand{\weird}{{\tt weirddetector}}
\newcommand{\cofiam}{{\tt CoFiAM}}
\newcommand{\phasma}{{\tt phasma}}
\newcommand{\github}{\href{https://github.com/joheenc/woi_data}{this URL}}

\title[New Periodic \TESS\ Signals]{
Hundreds of new periodic signals detected in the first year of \TESS\ with the \weird
}
\author[Chakraborty, Wheeler \& Kipping]{Joheen Chakraborty$^{1}$\thanks{E-mail:
\href{mailto:jc5110@columbia.edu}{jc5110@columbia.edu}}, Adam Wheeler$^{1}$ and David Kipping$^{1,2}$\\
$^{1}$Dept. of Astronomy, Columbia University, 550 W 120th Street, New York NY 10027\\
$^{2}$Flatiron Institute, 162 5th Av., New York, NY 10010}

\date{Accepted . Received ; in original form }

\pubyear{2020}

\begin{document}
\label{firstpage}
\pagerange{\pageref{firstpage}--\pageref{lastpage}}
\maketitle

\begin{abstract}
We apply the \weird, a nonparametric signal detection algorithm based on phase
dispersion minimization, \textbf{in a} search for low duty-cycle periodic signals in the
Transiting Exoplanet Survey Satellite (\TESS) photometry. \textbf{Our approach}, in contrast to
commonly used model-based approaches specifically for flagging transits,
eclipsing binaries, or other similarly periodic events, makes minimal
assumptions about the shape of a periodic signal, with the goal of finding
``weird'' signals of unexpected or arbitrary shape. In total, 248,301 \TESS\ sources
from the first-year Southern sky survey are run through the \weird, of which
we manually inspect the top 21,500 for periodicity. To minimize false-positives,
we here only report on the upper decile in terms of signal score, a sample for which
we obtain 97\% recall of \TESS\ eclipsing binaries and 62\% of the TOIs. In our
sample, we find 377 previously unreported periodic signals, for which we make a
first-pass assignment that 26 are ultra-short periods ($<0.3$\,d), 313 are likely
eclipsing binaries, 28 appear planet-like, and 10 are miscellaneous signals.
\end{abstract}


\section{Introduction}
\label{sec:intro}

With the recently-launched Transiting Exoplanet Survey Satellite (\TESS;
\citealt{ricker:2015}) observing on the order of $10^6$ unique targets across
the full sky and boasting an anticipated yield of thousands of novel exoplanets
\citep{sullivan:2015,sullivan:2017,bouma:2017,barclay:2018,ballard:2019}, there
is a need for varying approaches to the analysis of \TESS\ data products to make
full use of the survey's wide scope and potential. The large survey size and
all-sky nature of \TESS\ make it particularly adept for catching uncommon
signals, of potentially ill-classified or even unknown shape. These ``needles
in the haystack'' may challenge our understanding of other stars and spark
new theoretical developments, such as the case of Boyajian's Star
\citep{boyajian:2016,wright:2016,bodman:2016,foukal:2017,katz:2017,
nelusan:2017,ballesteros:2018,wyatt:2018,martinez:2019,sucerquia:2019}.
Accordingly, new methods to flag these signals for closer inspection have
been developed in recent years to expedite their recovery \citep{giles:2019,
schmidt:2019,wheeler:2019}.
 
The development of these \textbf{new methods} also serves to benefit the recovery of
more conventional signals that may have been missed by more standard
approaches, such as Box Least Squares \citep{kovacs:2002,kovacs:2006}. For
example, many of the known \Kepler\ single- and double- transiting systems
come from a myriad of independently developed search methods
\citep{schmitt:2017,uehara:2016,wang:2015,dfm:2016,kawahara:2019}, \textbf{including one specifically} designed to find unusual signals
\citep{wheeler:2019}.
 
\textbf{The latter approach}, known as the \weird, was previously successfully run on the
161,786 \Kepler\ sources, where it identified the 6 aforementioned missed
mono/stereo transiters, but also another 18 previously missed periodic signals
consistent with binary star systems. Given the similarities between the \Kepler\
and \TESS\ data products, this makes the \weird\ attractive as an
automated search tool for \TESS\ too. The \weird\ is a signal detection
algorithm with the goal of relaxing assumptions about signal shape in order to
detect events of abnormal morphology \citep{wheeler:2019}. Whilst the \weird\
is \textbf{not specifically intended for exoplanet transit detection}, it is related to
the more planet-specific search methods, in that it accounts for a different
set of potential signals. Accordingly, the \weird\ is able to detect a wider
variety of signals than just, say, transiting planets or eclipsing binaries (though it is also able to recover to those events),
at the cost of lower sensitivity to signals of \textit{a-priori} known shape.
   
In this work, we \textbf{apply the \weird\ to} 248,301 target light curves from
sectors 1 through 13, which represents \TESS's first-year survey of the Southern sky. From these, \textbf{we manually identify}
468 significant periodic signals which appear to not have been previously reported, with particular attention given to the top 30 of these. In
Section~\ref{sec:methods}, we briefly describe the algorithm and the specific
changes made for \TESS. In Section~\ref{sec:results}, we discuss the
application of the code, cross-reference the findings with those of other
catalogues, and introduce the most likely signals detected which are previously
undetected at the time of writing. In Section~\ref{sec:discussion}, we discuss
further improvements and potential continuations of our work.

\section{Methodology}
\label{sec:methods}

The basic principle of the \weird\ is to \textbf{identify high-likelihood periodic signals for manual inspection} while abandoning
any parametric function to describe the transit shape. \textbf{We emphasize that the \weird\ algorithm cannot identify signals all on its own, but instead facilitates a subsequent informed manual search through the high-probability signals sorted by the algorithm; as such, it is most useful to consider usage of the \weird\ algorithm as one part in a larger search pipeline}. A more in-depth
discussion of the \weird\ is available in \citet{wheeler:2019}, but we shall also include a brief
description, highlighting in particular the points specific to our
application on \TESS\ data.

\subsection{General principles of \weird}

As the \weird\ searches for coherent, low duty-cycle periodic signals, the
algorithm folds on many trial periods and searches for the period values which
exhibit the smallest dispersion of fluxes. We note that phase dispersion minimization is not a novel concept--it has been used in other algorithms such as the Plavchan periodogram \citep{plavchan:2008,parks:2014}. In the case of \TESS, which monitors
each field for ${\simeq}27$ days, we opt for a period grid ranging from
0.25\,days up to 15\,days uniform in $\log P$ (24,567 total period values).
Just as was found with \Kepler\ though \citep{wheeler:2019}, the algorithm is
sensitive to periods outside the period grid if there is a harmonic of the true
period within the grid (which is the case for some of our flagged signals
later).
    
To quantify the relative success at phase dispersion minimization of
a given trial period, the \weird\ examines the chi-squared ($\chi^{2}$,
which reflects how scattered the values are) and kurtosis ($\kappa$, which
reflects the weight of a distribution's tail relative to its peak). The
algorithm then calculates \textit{excess} kurtosis, $\kappa' = \kappa - 
\kappa_{\mathrm{Gaussian}}$ and chi-squared less than the median, $\Delta\chi^2=\mathrm{median}(\chi^2)-\chi^2_\mathrm{period}$. These
two metrics are multiplied to form a score strength, such that we require
both perform well to yield a putative signal. To normalize the scores,
we re-calculate $\kappa'(P)\Delta\chi^2(P)$ along a grid of periods but
for a scrambled version of the original data (which should have no periodicity),
and evaluate a sliding standard deviation of the result, $\sigma(P)$.
Using these quantities, the algorithm defines a final merit function, $\zeta$, for
each period value, given by $\zeta(P) = \kappa'(P)\Delta\chi^2(P)/\sigma(P)$.
    
The above value of the merit function gives us intuition about the ideal signal the \weird\
is searching for. To maximize the merit function of a period
$P_{\mathrm{ideal}}$, the phase curve should be much more coherent at
$P_{\mathrm{ideal}}$ than at other ``random" values of $P$ (i.e. large
decrease in $\chi^2$); the phase curve should hover around a constant value
with only one (or a few) short excursion(s) from the baseline (i.e. high excess
kurtosis $\kappa'$); and these two conditions should otherwise have low
variation at nearby $P$-values not equal to $P_{\mathrm{ideal}}$, maintaining
small values for $\Delta\chi^2$ and $\kappa'$ to demonstrate there is, in fact, a distinct signal with a specific period more likely than the rest (i.e. low $\sigma(P)$).
    
\subsection{Differences between \Kepler\ and \TESS}
    
The \weird\ is highly versatile and generalizable, and was able to run on \TESS\ data after appropriate changes were made from the implementation for \Kepler.
Following are the details that are sensitive to the particular
telescope gathering the data.
    
\Kepler\ observed a 115 square-degree patch of sky, at a cadence of 30 minutes
for 17 quarters of up to 90 days each. \TESS, on the other hand, generates
light curves with a cadence of 2 minutes and will observe 26 different sectors
for ${\simeq}$27 days each. \TESS, while it spends only a month on the targets of each
sector (barring targets that receive extra coverage as a result of overlap
between sectors), will survey 85\% of the sky, covering roughly 400 times as
many unique stars as \Kepler. As a result, there will be far more data to
search through; however, by virtue of the shorter baseline, we are more limited in the range of periods for which we can detect
signals.
    
In addition, having fewer data points means \textbf{the \weird\ will be} more sensitive to outlier events when calculating both $\chi^2$ and $\kappa$, such
as stellar flares or instrument-caused deviations. As a result, we were more
stringent in our outlier rejection when considering the phase-folding part of
the algorithm to limit one-time events from inducing spurious signal detections
or leading to false positives by unfavorably skewing $\kappa$. Additionally,
while we do not expect $\kappa$ or $\sigma$ to change much, $\chi^2$ (and
therefore $\Delta\chi^2$) is orders of magnitude lower in \TESS\ because of the
fewer points in the light curves. This means that $\zeta$ drops accordingly;
however, because the \emph{absolute} values of $\zeta$ do not matter as much as
the \emph{relative} values for determining the most promising period values, and we take the step of dividing by $\sigma(P)$, our analysis is little-affected.
    
We also had to adjust the bandwidth used for our median-filter detrending
algorithm to better reflect the duration of the signals we were searching for
while considering the shorter range of trial periods we were folding on. In
addition, the presence of \TESS's regular data gaps posed an issue, as \Kepler\
data suffered far fewer gaps per unit time. However, as these changes did not require fundamentally altering the algorithm, their treatment is
discussed in the beginning of Section~\ref{sec:results}.
        
\subsection{Aliases}

One major limitation of \textbf{the \weird\ is} its tendency to flag rational
fractions or multiples (``aliases") of the true period of a signal. This 
is unavoidable when using a folding technique with such relaxed 
requirements for signal shape, and the issue is not specific to \TESS\
light curves.
    
Here, we partially corrected for this error by attempting to automatically flag
the correct period when an integer multiple was identified in the case of
single-dip signals. It is important to note that a) our technique \emph{only}
accounts for flagged periods that are an integer multiple of the true period
(which we chose to target because it is the most common case of aliasing); and
b) we still need to manually examine the data to determine if the period
detected is truly the correct one. Still, our technique helps us identify
signals by suggesting the correct period more frequently.
    
\begin{figure*}
\centering
\includegraphics[width=\hsize]{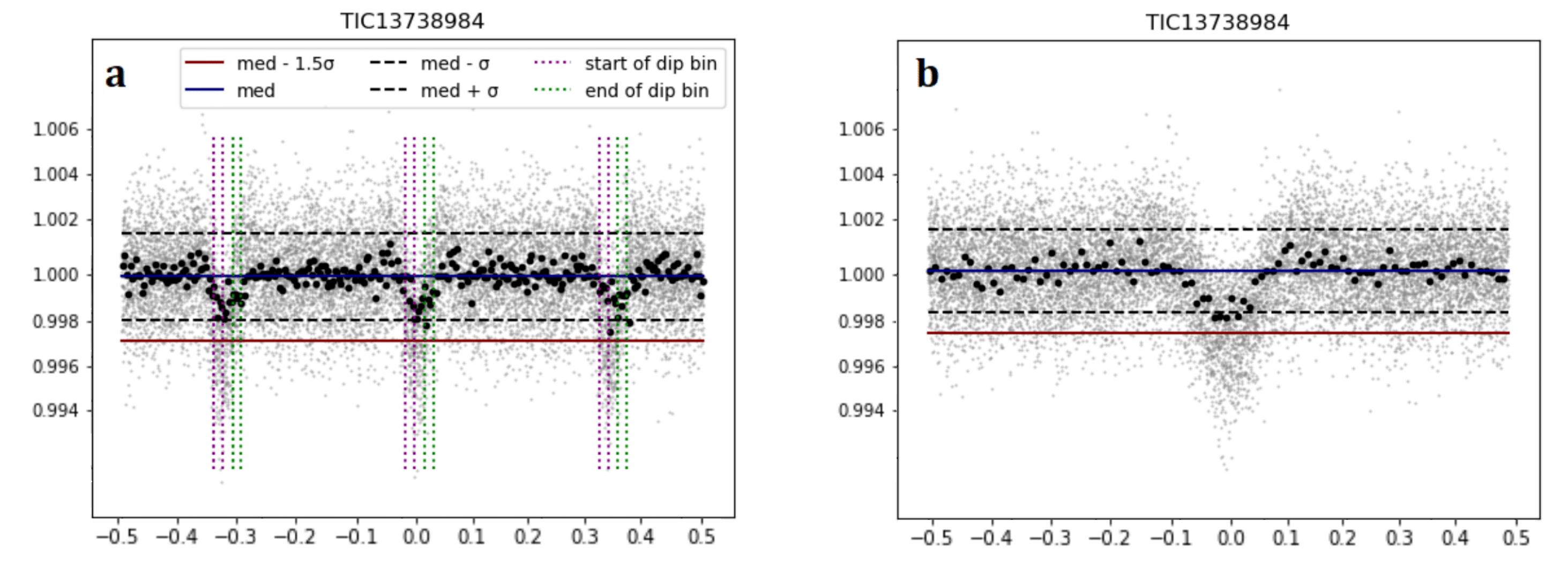}
\caption{
\textbf{a:} A sample flagged signal (TIC 13738984) with $\zeta_{\mathrm{peak}}$
at a $P_{\mathrm{flagged}}$ which is an integer multiple of
$P_{\mathrm{true}}$. Here, $\frac{P_{\mathrm{flagged}}}{P_{\mathrm{true}}}=3$.
The solid blue line indicates median($F$); the solid maroon line indicates
median($F$) $-1.5\sigma$, the threshold below which we take the beginning of a
dip; and the dashed gray lines indicate median($F$) $\pm\sigma$, the range at
which we take the end of a dip. The vertical lines indicate the boundaries of
the 200-point bins which mark the starts and ends of the dips. \textbf{b:} The
result of using the algorithm, which successfully picks out the correct period.
}
\label{FigVibStab}
\end{figure*}

For a single-dip event with true period $P_{\mathrm{true}}$ and flagged period
$P_{\mathrm{flagged}}$, there will be
$\;n\;=(P_{\mathrm{flagged}}/P_{\mathrm{true}})\;:\;n\;\in\;\mathbb{Z}^+$
dips in the phase curve folded with $P_{\mathrm{flagged}}$. By automatically
detecting the number of dips $n$, we can therefore retrieve the correct period
by multiplying. We identify a dip by the following: examine the points in
folded time-order in 200-point sliding windows. If greater than one-third
of the points in this window have a flux value that falls under $1.5\sigma$
below the median, note the window as the beginning of a dip. Then, wait for
the flux values to return to normal before looking for another dip by flagging
the \emph{end} of a dip when greater than three-quarters of the window has a
value within $\sigma$ of the median (see Figure~\ref{FigVibStab}). Note that the specific time
picked out by the sliding window is not important, as this approach does not
need to consider the duration or shape or the dip--only the \emph{number} of significant dips $n$.
    
The specific numerical values were chosen heuristically to handle the
aforementioned case of aliasing-- by far the most common case within our data.
Though a similar approach may work in principle for regular increases in flux,
there are too many confounding factors (particularly flares) to have a reliable
solution to that problem.
    
\section{Application to \TESS\ data}
\label{sec:results}

\subsection{Data preparation}

We ran the \weird\ on all light curve files in sectors 1 through 13 available at
\href{http://archive.stsci.edu/tess/bulk_downloads/bulk_downloads_ffi-tp-lc-dv.html}{\color{blue}{MAST}}
(248,301 not including overlap of targets between sectors;
223,087 unique targets after considering overlap). \textbf{We remove probable outliers
from each PDC light curve by removing points more than $6*\mathrm{MAD}$
away from the median in an 11-point rolling window, where
$\mathrm{MAD}$ is the median absolute deviation: $\mathrm{median}(|F_i-\mathrm{median(\textbf{F})}|)$. (As one of the key assumptions of \weird\ is the presence of Gaussian noise, we can use $\mathrm{MAD}$ as an estimator of the standard deviation with $\sigma = 1.4826*\mathrm{MAD}$, meaning our threshold is also roughly equal to $4\sigma$)}. We also removed all points with non-zero quality flags in the FITS files.
    
In each sector, we split each PDC time series into two segments (``semi-sectors'')
before and after the ${\sim}1$ day data gap. Within each segment, we used linear
interpolation to fill in missing data and detrended with a moving median
filter, using a 1\,day bandwidth, to remove long-term trends, following
\citep{wheeler:2019}. We decided on the 1-day window as it is both sufficiently
shorter than the duration of observation (27 days per target) and longer than
the duration of our signals of interest (most are of order
$10^{-1}$ days). This also acts to suppress long duty-cycle signals associated with stellar rotation.
After detrending, we removed the interpolated points and
recombined the segments. \textbf{The \weird\ was then run} on the
$24,567$ trial periods for 248,301 unique light curves.
   
\subsection{Filtering}
   
In order to preserve only the most likely signal candidates and remove artifacts of the data analysis, some cuts were made to the resultant data. The first one we note is that upon
examining a scatter plot of a light curve's $\log \zeta_{\mathrm{peak}}$ versus $\log P$,
one immediately sees piling up a certain frequencies. A similar effect was
reported by \citet{wheeler:2019} on the \Kepler\ data. These are immediately
suspicious as spurious common modes, rather than astrophysical signals. We
therefore proceeded to design a means to filter out these suspicious periods.

\begin{figure*}
\centering
\includegraphics[width=\textwidth]{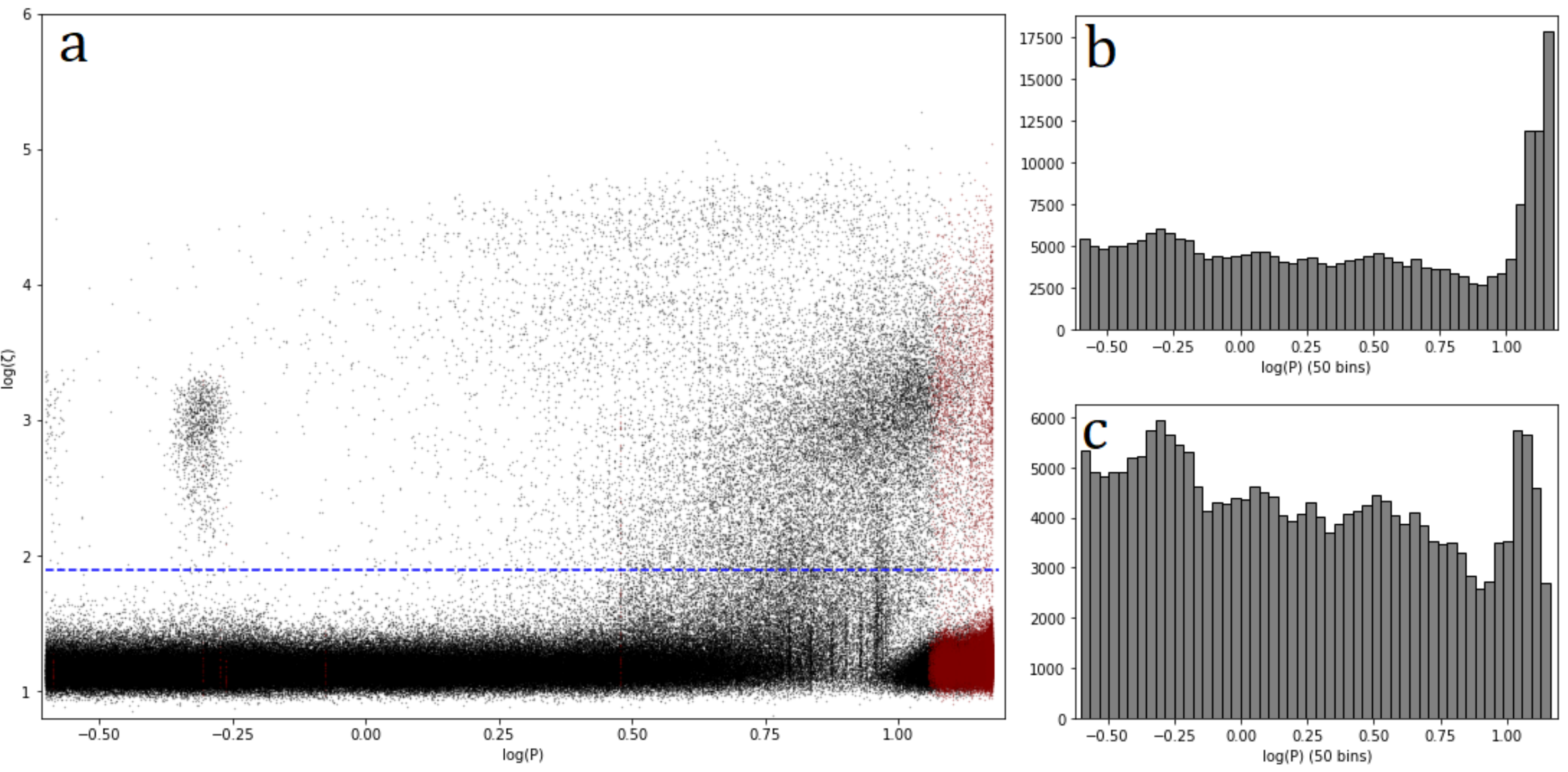}
\caption{
\textbf{a:} Flagged period values with cut (over-represented) periods marked in
red. The blue line represents $\zeta_{\mathrm{threshold}}$, the cutoff for
$\log_{10}(\zeta_{\mathrm{peak}})$ above which we manually inspected for signals.
\textbf{b:} Histogram of $\log_{10}(P)$ before cutting over-represented period
values (50 bins). There is significant saturation towards the tail end, as
represented in Fig. 2a. \textbf{c:} Histogram of $\log_{10}(P)$ after cutting
over-represented periods (50 bins). This graph is more evenly distributed,
as desired.
}
\label{fig:VibStab}
\end{figure*}

Since the distribution of periods guessed by the \weird\ was at first not log-uniform (in particular with large saturation towards the upper end), we cut the 1,000 most commonly appearing periods to achieve a more uniform distribution (Figures~\ref{fig:VibStab}b-c). Though the precise number chosen as a cutoff was somewhat ad-hoc, it accomplished the desired goal of picking out a set of signals with a more uniform distribution of $\log(P)$.

\textbf{Even after cuts, there are a few peculiarities in the distribution worth addressing. There is a low density of points at $\log(P)\approx 0.8$; we posit these are caused by phasing of periods which are almost exactly half ($\approx6.5$) of the 13-day semi-sector baseline, causing dips to frequently fall within the one-day gap.} We also note a concentration of high-$\zeta_{\mathrm{peak}}$ points close to $\log(P)=-0.3$; these could be caused by a common mode, as they are unlikely to be astrophysical considering their distance from the rest of the distribution. \textbf{The high-$\zeta$ signals are denser at the higher end of $P$-values even after cuts, which is likely due to lower $\chi^2$ values caused by the lesser dispersions associated with folding on longer periods; this trend contributes significantly to the aliasing effect of rational multiples $P_{\mathrm{true}}$ as noted in  Section 2.3}

Since our goal is to find new signals, known variables from the Villanova eclipsing binary catalogue \citep{prsa:2020}, the ASAS-SN catalog of variable stars \citep{shappee:2014, jayasinghe:2018, jayasinghe:2019}, the International Variable Stars Index (VSX, \citet{watson:2006}), and the Tess Objects of Interest (TOIs) (\href{https://exoplanetarchive.ipac.caltech.edu/cgi-bin/TblView/nph-tblView?app=ExoTbls&config=TOI}{NASA Exoplanet Archive list of current TOIs}) were also removed. Altogether, these cuts comprised roughly 13\% of the data, leaving 215,871 of the 248,301 targets intact.

We consider a signal to be ``interesting'' if its periodogram has a
most-significant peak such that $\zeta\geq\zeta_{\mathrm{threshold}}$,
again following \citet{wheeler:2019}. We choose $\zeta_{\mathrm{threshold}}$
not independently at some fixed value, but set it to pick
out the top decile of candidates, after cutting both common periods and already-discovered signals, for manual
inspection. We look only at the most-significant peak (and rational fractions)
as found by the previously described method, to generate phase curves with the
highest likelihood of containing the correct period.

\subsection{Data products}

After filtering, we flag 377 previously-unidentified targets (dubbed Weird
Objects of Interest, or WOIs) that appear to contain some periodic signal.
The signals were \textbf{selected manually based on} whether their phase curves showed a clear signal (or multiple clear signals, in the case of aliases). $\zeta_{\mathrm{peak}}$ values of the WOIs fall within the range of $10^3$ to $10^5$, providing evidence that the selected periods are several orders of magnitude better than naive guessing. Additionally, we examined distributions of
$\Delta\chi^2$ and $\kappa$ for the WOIs, which showed no dependence on each
other, assuring us that the high $\zeta_{\mathrm{peak}}$ values are not
artifacts. In Table~\ref{tab:WOIs}, we present the $\zeta_{\mathrm{peak}}$ and periods, along with accompanying information, for
the WOI signals reported in this work.

\begin{figure*}
\centering
\includegraphics[width=1\textwidth]{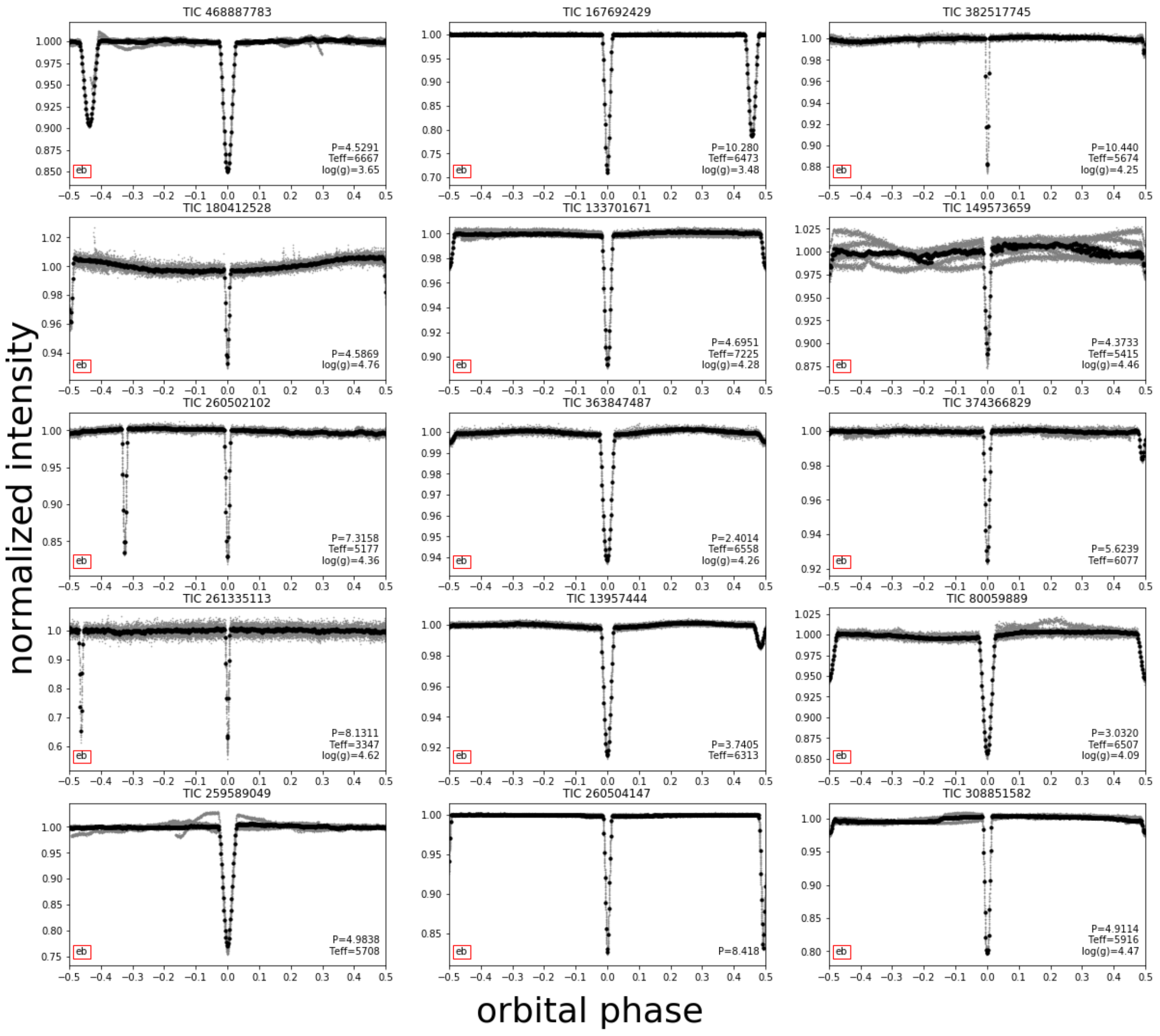}
\caption{
The first to fifteenth highest scoring novel signals found 
by the \weird\ in this work (in decreasing order of
$\zeta_{\mathrm{peak}})$.
}
\label{fig:WOIs1}
\end{figure*}

\begin{figure*}
\centering
\includegraphics[width=1\textwidth]{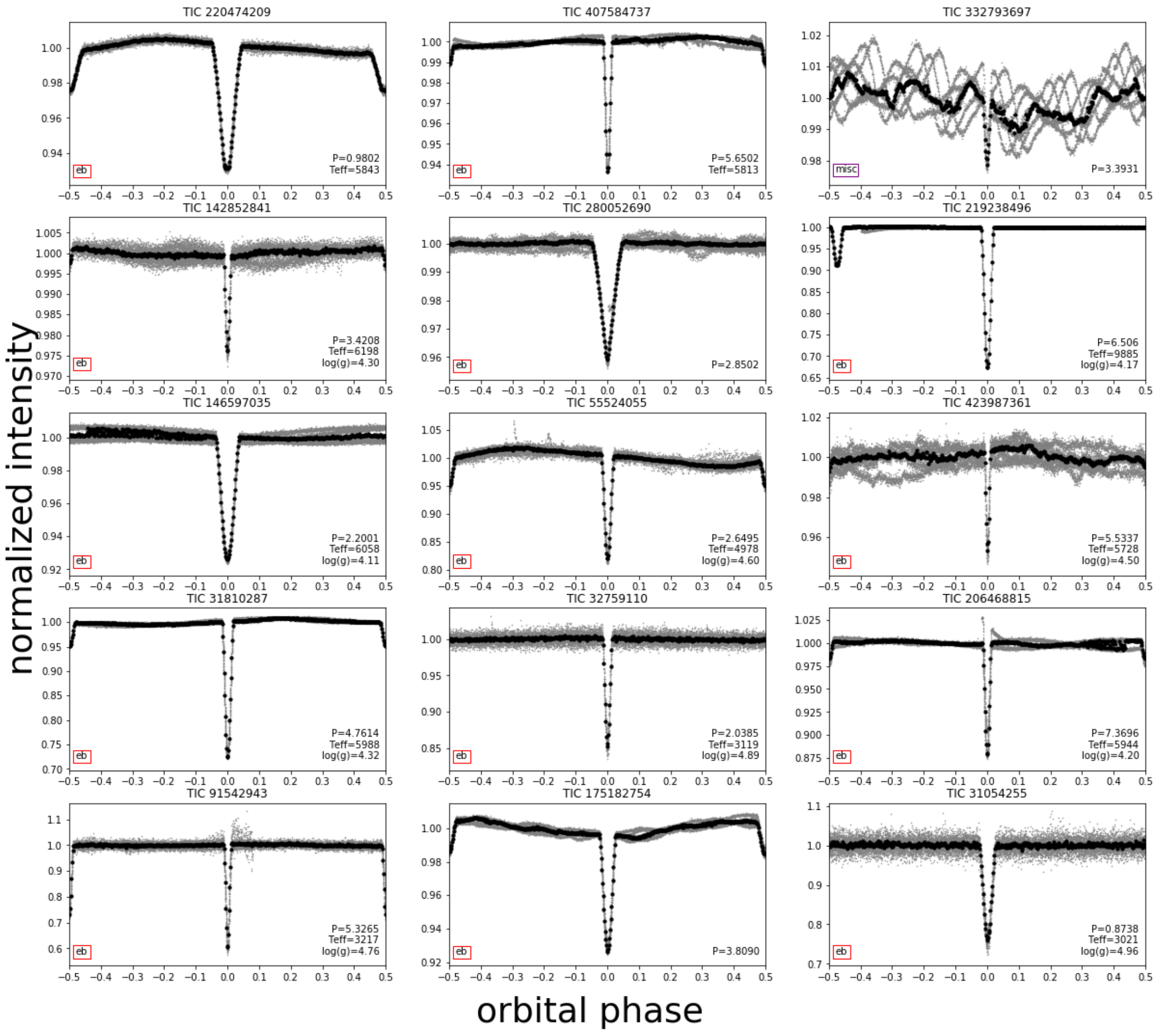}
\caption{
The sixteenth to thirtieth highest scoring novel signals found 
by the \weird\ in this work (in decreasing order of
$\zeta_{\mathrm{peak}})$.
}
\label{fig:WOIs2}
\end{figure*}

To ensure each signal was not an artifact of the median-filter detrending, we
applied different detrending algorithms to the raw light curves and re-folded
on the flagged period. We applied the implementation of Cosine Filtering with Autocorrelation Minimization (\cofiam; \citealt{kipping:2013}) found in the
\href{https://github.com/alexteachey/MoonPy}{\color{blue}{MoonPy Python
library}}, as well as \phasma, a non-parametric phase-folding algorithm
\citep{jansen:2018}. Example phase curves presented in Figures~\ref{fig:WOIs1} 
\& \ref{fig:WOIs2} have been detrended with \phasma. We also queried the
TIC-8 catalog \citep{stassun:2018} for these objects to report their basic
stellar properties, which is provided in the figure panels where available.
We provide moving
median, \cofiam\ and \phasma\ detrended phase curves for the 337 signals, as well as a CSV form of Table~\ref{tab:WOIs} at \github.

\subsection{Performance}

\begin{figure*}
\centering
\includegraphics[width=\textwidth]{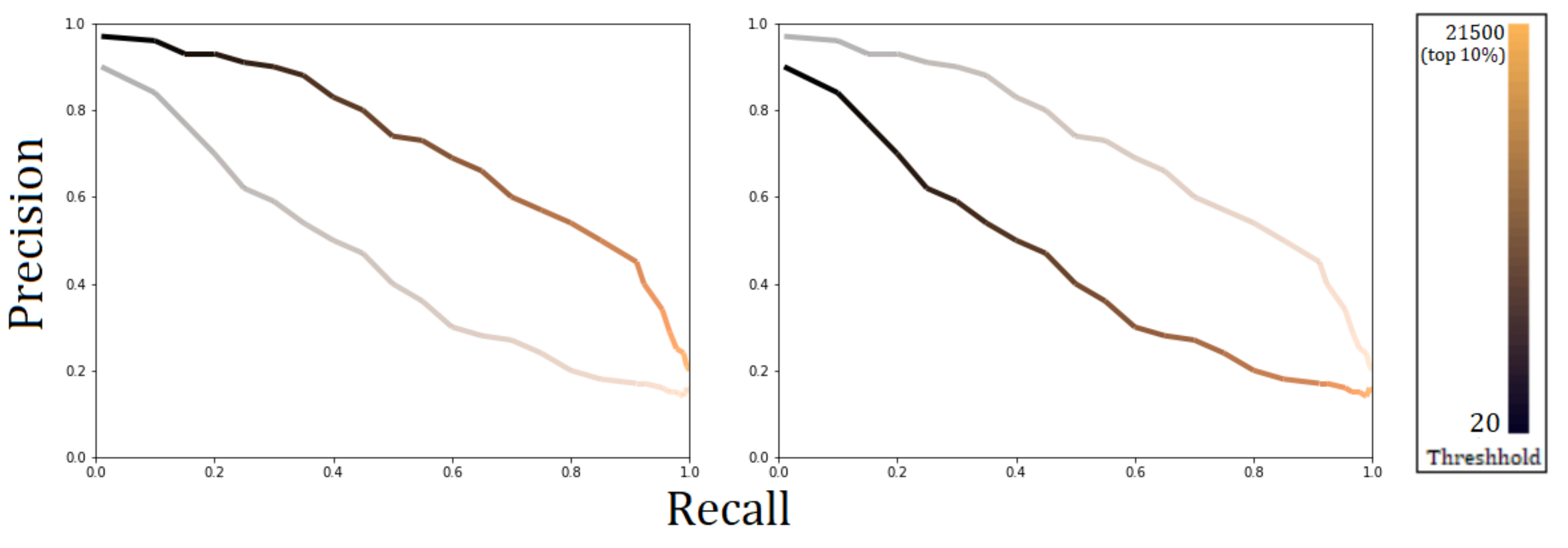}
\caption{
Precision-recall curve for our algorithm applied to \TESS\ data, with
(\textbf{a}) and without (\textbf{b}) cuts. To generate different points, we
vary the total number of signals we consider (i.e. the percentile of
$\zeta_{\mathrm{peak}}$ we take for manual triage). To define precision and
recall, we take true positives as signals documented in the Villanova eclipsing
binaries and \TESS\ Objects of Interest, as well as visible signals that
passed our data-artifact sanity checks; and false positives as flagged signals
not meeting those requirements (most commonly just phase curves with background
noise).}
\label{fig:recall}
\end{figure*}

To quantify the \weird's performance, we analyzed its effectiveness at
recovering signals from the Villanova eclipsing binary and TOI catalogues. At the time of writing, 62\% of currently-documented TOIs
(1267/2044) and 97\% (1801/1855) of Villanova eclipsing binaries fell within
the top decile of $\zeta_{\mathrm{peak}}$ values. The missing targets in each
catalog are expected, as ideally a model-based approach would be best for
finding transiting planets and eclipsing binaries, respectively, whereas 
the \weird\ is sensitive to all manner of periodic signals. \textbf{We note also that all of the missing targets from the Villanova EB catalogue have periods falling outside our period grid ($>15$ days), explaining the algorithm's failure to recover them.} We maintained a
high recall (the fraction of catalogued true signals recovered) percentage at
the cost of precision (the fraction of flagged signals which we know or believe
are real), as we wanted to minimize false negatives over
false positives to find as many interesting signals as possible. Of course, one
could vary the fraction of signals we consider from 10\% to modify the
precision and recall percentages (Figure~\ref{fig:recall}).
    
\section{Discussion}
\label{sec:discussion}

In this work, we have \textbf{applied the \weird\, designed to flag high-likelihood periodic
signals of arbitrary shape in time series photometry for subsequent manual inspection}, to 248,301 unique \TESS\
targets in Sectors 1-13 (the year-one Southern sky survey). Full frame image targets were not analyzed. Our work
identifies 377 very high signal to noise periodic signals that have not been
previously flagged in the literature or elsewhere to our knowledge. We have
provided basic parameters and identifiers for these systems to enable
community follow-up.
The number of WOIs could surely be increased by relaxing our threshold to
consider more than the top decile of $\zeta_{\mathrm{peak}}$ events. This choice
was made to allow us to quickly hone in on the most significant signals - 
the goal is to perform triage here and simply see if there is something
glaringly obvious and exciting in the \TESS\ data. Having done so, we
can report that amongst the 377 WOIs identified, none of them appear to
display the kind of remarkably unusual behaviour observed for Boyajian's
Star \citep{boyajian:2016}, despite the fact the \weird\ easily recovered
that signal in the \Kepler\ timeseries\footnote{Note that Boyajian's Star
is not strictly periodic, but like the detection of mono-transits by the
\weird, the kurtosis aspect of its merit function provides sensitivity
to even one-off events - although it is certainly not optimized to them.}.
This does not mean analogs do not exist amongst the target stars. \textbf{While the new signals uncovered in this survey suggest gaps in the completeness functions of other detection methods, it remains unclear exactly what these gaps are. One could go about precisely identifying them with a detailed injection and recovery hare-and-hound exercise across several different search strategies, but that is beyond the scope
of this work. By the same token, we cannot directly convert our lack of detection into an occurrence rate of Boyajian-like stars (nor is it immediately clear how one would define such events).}

We also highlight that it is not the objective of this work to offer
astrophysical interpretations of the identified signals. They are simply
flagged periodic signals that have been apparently missed thus far, in the
same vein as the \Kepler\ study by \citet{wheeler:2019}. Nevertheless, we will
offer some simple classifications below.

We highlight that the periods in Table~\ref{tab:WOIs} are only estimates
and it is possible the true period is an alias. Still, in many
cases we find extremely short flagged periods, less than 0.3\, days, which we
classify as ``ultra short period'' (USP). If correct, these
are unlikely to be eclipses since for a Solar-like host the orbital period
would be grazing or inside the star. A possible explanation is a rapidly
rotating star, which could be checked with a $v \sin i$ measurement, or
pulsations - which might be expected in the case of evolved stars with
lower $\log g$ values. As an example, TIC 443494351 is a USP but also has an 
excessively long duty cycle for its ``dip'' (which is itself questionable
in nature). This makes it unlikely to be an eclipse signal and more
consistent with intrinsic stellar activity. We classify 26 of the cases as
USPs.

The overwhelming majority of  WOIs appear to be eclipsing binary systems, as evident from
inspection of Figures~\ref{fig:WOIs1} \& \ref{fig:WOIs2}. We can crudely
classify an ``EB'' as one which displays a) two dips per period, or b)
one dip and noticeable phase curve variations, or c) one dip with a depth
greater than 2\%. Using this criterion, and excluding the ultra short
periods mentioned above, we classify 313 of the WOIs as likely EBs.

A third classification we offer is ``planet-like'' (PL). These are signals
which appear consistent with a transiting planet from the phase curve
morphology alone, but have not been subject to any further
vetting. We classify signals as planet-like if they show no noticeable
phase curve variations and have one-dip per period which is less than
2\% in depth. We also impose the condition that the duration of the
dip must be less than one-sixth of the orbital period (which corresponds
to a planet with $a/R_{\star}>2$). This results in 28 planet-like signals.
This leaves 10 WOIs without a clear classification - within which we 
suspect stellar activity may be broadly responsible.

Amongst the WOIs, there are some curious signals, such as TIC 24347173.
A 2\% phase curve appears consistent with rotational modulation giving
the phasing with respect to the dips, yet the strict coherence implies
an apparent 2:1 commensurability between the rotation and orbital period.
\citet{beky:2014} proposes an explanation to signals of this nature: if there
is a latitude on the surface of a star with a period synchronous to its
satellite, magnetic interactions between the objects may result in preferential
spot formation at that latitude. The star would then exhibit photometric
variations in resonance with its companion.

TIC 197886566 bares some resemblance to this but the phase curve is
perhaps more plausibly consistent here with being caused by the
companion, although this would require an offset in the phase curve.

Our work provides the first demonstration of an en-masse search for weird signals within the \TESS\ photometric data. Although no clear examples of highly irregular signals manifest, the fact we recover 313 EB-like signals and 28 planet-like signals suggests gaps in the completeness functions of standard algorithms and demonstrates the utility of numerous algorithms being applied to photometric data sets such as this
	
\section*{Acknowledgments}

DMK acknowledges support from the Simons Foundation and the Alfred P. Sloan Foundation.
Special thanks to Tom Widdowson, Mark Sloan, Laura Sanborn, Douglas Daughaday, Andrew
Jones, Jason Allen, Marc Lijoi, Elena West, Tristan Zajonc, Chuck Wolfred, Lasse Skov \& Martin Kroebel.

\section*{Data Availability}

The data underlying this article are publicly available at \href{https://github.com/joheenc/woi_data}{https://github.com/joheenc/woi\_data}.

\clearpage
\onecolumn
\appendix
\section*{Appendix}

\begin{longtable}{lcccccccccccc} 
\caption{\emph{Parameters for all 377 WOIs (grouped by classification and in descending order by $\zeta_\mathrm{peak}$.})}\\
\label{tab:WOIs}\\ 

\hline\hline
TIC ID & $P$ [d] & $\zeta_{\mathrm{peak}}$ & $\Delta\chi^2$ & $\kappa$ &  $T_{\mathrm{eff}}$ [K] & $\log g$ & dips & $\delta$ [\%] & RA & Dec & Flat baseline & Class\\
\hline
\endhead
\hline
\endfoot

468887783	&	4.5291	&	1.16e+05	&	8.20e+03	&	1.19e+02	&	6667	&	3.65	&	2	&	14.992	&	56.67	&	9.90	&	y	&	eb\\
167692429	&	10.280	&	1.07e+05	&	1.51e+03	&	1.63e+02	&	6473	&	3.48	&	2	&	28.962	&	102.71	&	-63.43	&	y	&	eb\\
382517745	&	10.440	&	1.00e+05	&	4.85e+02	&	1.82e+02	&	5674	&	4.25	&	2	&	12.356	&	117.45	&	-64.04	&	y	&	eb\\
180412528	&	4.5869	&	9.97e+04	&	6.50e+03	&	8.81e+01	&	---	    &	4.76	&	2	&	6.7849	&	11.35	&	-77.86	&	y	&	eb\\
133701671	&	4.6951	&	9.55e+04	&	1.55e+04	&	1.42e+02	&	7225	&	4.28	&	2	&	10.686	&	229.68	&	-48.90	&	y	&	eb\\
149573659	&	4.3733	&	9.47e+04	&	3.70e+03	&	1.27e+02	&	5415	&	4.46	&	1	&	11.673	&	86.61	&	-61.69	&	n	&	eb\\
260502102	&	7.3158	&	9.37e+04	&	6.17e+03	&	1.02e+02	&	5177	&	4.36	&	2	&	17.428	&	96.51	&	-59.01	&	y	&	eb\\
363847487	&	2.4014	&	8.05e+04	&	1.12e+04	&	5.53e+01	&	6558	&	4.26	&	1	&	6.1775	&	200.43	&	-42.62	&	y	&	eb\\
374366829	&	5.6239	&	7.86e+04	&	2.67e+03	&	1.46e+02	&	6077	&	--- 	&	2	&	7.6074	&	162.71	&	0.52	&	y	&	eb\\
261335113	&	8.1311	&	7.85e+04	&	1.45e+03	&	1.53e+02	&	3347	&	4.62	&	2	&	40.559	&	92.33	&	-80.69	&	y	&	eb\\
13957444	&	3.7405	&	7.82e+04	&	2.45e+04	&	1.00e+02	&	6313	&	--- 	&	1	&	8.4917	&	75.48	&	-37.03	&	y	&	eb\\
80059889	&	3.0320	&	7.74e+04	&	1.35e+04	&	7.07e+01	&	6507	&	4.09	&	2	&	14.450	&	297.89	&	-43.37	&	y	&	eb\\
259589049	&	4.9838	&	7.67e+04	&	2.93e+03	&	1.37e+02	&	5708	&	--- 	&	1	&	22.989	&	72.43	&	-53.00	&	y	&	eb\\
260504147	&	8.418	&	7.43e+04	&	5.35e+04	&	1.12e+02	&	---	    &	---	    &	1	&	17.577	&	96.43	&	-55.16	&	y	&	eb\\
308851582	&	4.9114	&	7.05e+04	&	1.09e+03	&	1.18e+02	&	5916	&	4.47	&	2	&	20.297	&	122.58	&	-63.74	&	y	&	eb\\
220474209	&	0.9802	&	6.84e+04	&	1.22e+05	&	3.09e+01	&	5843	&	---	    &	2	&	6.9241	&	75.38	&	-59.51	&	y	&	eb\\
407584737	&	5.6502	&	6.77e+04	&	1.80e+04	&	1.60e+02	&	5813	&	--- 	&	2	&	6.3343	&	352.77	&	-74.13	&	y	&	eb\\
332793697	&	3.3931	&	6.64e+04	&	4.16e+03	&	5.97e+01	&	--- 	&	--- 	&	1	&	2.0910	&	215.98	&	-50.49	&	n	&	misc\\
142852841	&	3.4208	&	6.54e+04	&	4.29e+03	&	8.79e+01	&	6198	&	4.30	&	2	&	2.4215	&	50.50	&	-31.10	&	y	&	eb\\
280052690	&	2.8502	&	6.52e+04	&	4.10e+03	&	9.08e+01	&	---	    &	---	    &	1	&	4.0962	&	45.44	&	-76.20	&	y	&	eb\\
219238496	&	6.506	&	6.49e+04	&	8.86e+03	&	2.06e+02	&	9885	&	4.17	&	2	&	32.689	&	64.49	&	-48.39	&	y	&	eb\\
146597035	&	2.2001	&	6.37e+04	&	1.34e+04	&	8.37e+01	&	6058	&	4.11	&	1	&	7.3864	&	77.77	&	-21.49	&	y	&	eb\\
55524055	&	2.6495	&	6.34e+04	&	1.16e+04	&	7.65e+01	&	4978	&	4.60	&	2	&	18.112	&	72.05	&	-63.69	&	y	&	eb\\
423987361	&	5.5337	&	6.20e+04	&	3.27e+03	&	1.36e+02	&	5728	&	4.50	&	1	&	4.6543	&	235.84	&	-63.66	&	n	&	eb\\
31810287	&	4.7614	&	5.85e+04	&	3.13e+04	&	1.21e+02	&	5988	&	4.32	&	2	&	27.721	&	136.65	&	-47.06	&	y	&	eb\\
32759110	&	2.0385	&	5.79e+04	&	1.55e+04	&	8.20e+01	&	3119	&	4.89	&	1	&	14.696	&	86.24	&	-24.93	&	y	&	eb\\
206468815	&	7.3696	&	5.77e+04	&	2.61e+03	&	1.43e+02	&	5944	&	4.20	&	2	&	12.202	&	57.01	&	-53.60	&	y	&	eb\\
91542943	&	5.3265	&	5.67e+04	&	5.11e+03	&	1.05e+02	&	3217	&	4.76	&	1	&	40.042	&	43.82	&	-33.74	&	y	&	eb\\
175182754	&	3.8090	&	5.65e+04	&	6.71e+03	&	1.08e+02	&	---	    &	--- 	&	2	&	7.3807	&	115.96	&	-37.72	&	y	&	eb\\
31054255	&	0.8738	&	5.52e+04	&	1.45e+04	&	2.74e+01	&	3021	&	4.96	&	1	&	24.747	&	80.57	&	-25.11	&	y	&	eb\\
178938417	&	5.9747	&	5.48e+04	&	6.48e+03	&	1.98e+02	&	5405	&	4.51	&	2	&	0.8074	&	68.27	&	-24.01	&	n	&	eb\\
142409400	&	1.3869	&	5.42e+04	&	1.74e+04	&	2.59e+01	&	3385	&	4.58	&	2	&	19.947	&	103.09	&	-12.96	&	y	&	eb\\
178284729	&	2.2360	&	5.41e+04	&	2.78e+03	&	7.51e+01	&	3968	&	4.63	&	1	&	11.031	&	60.38	&	-20.45	&	y	&	eb\\
32150630	&	1.4921	&	5.38e+04	&	1.84e+04	&	4.36e+01	&	4323	&	--- 	&	2	&	14.503	&	59.24	&	-68.62	&	n	&	eb\\
143924219	&	2.8199	&	5.35e+04	&	4.54e+03	&	7.65e+01	&	6412	&	3.99	&	1	&	2.5442	&	332.70	&	-48.89	&	n	&	eb\\
339607421	&	2.4381	&	5.25e+04	&	1.33e+04	&	8.03e+01	&	6327	&	4.33	&	1	&	5.4663	&	46.20	&	-52.55	&	y	&	eb\\
238162238	&	1.2186	&	5.10e+04	&	3.65e+04	&	4.54e+01	&	6172	&	4.14	&	2	&	4.2458	&	106.39	&	-48.84	&	n	&	eb\\
421285598	&	2.3578	&	5.07e+04	&	4.77e+03	&	4.61e+01	&	5745	&	4.15	&	2	&	13.222	&	261.92	&	-54.11	&	n	&	eb\\
49534600	&	7.9529	&	5.06e+04	&	2.42e+03	&	1.52e+02	&	---	    &	---	    &	2	&	15.235	&	273.69	&	-52.81	&	y	&	eb\\
301956407	&	1.8314	&	4.99e+04	&	2.86e+04	&	5.52e+01	&	6079	&	4.09	&	2	&	2.6056	&	260.05	&	-70.04	&	y	&	eb\\
423530755	&	2.3421	&	4.97e+04	&	3.41e+04	&	8.64e+01	&	5547	&	4.27	&	1	&	7.2387	&	185.30	&	-20.70	&	n	&	eb\\
365694569	&	9.0344	&	4.88e+04	&	3.69e+04	&	1.13e+02	&	6358	&	3.99	&	2	&	34.617	&	81.63	&	9.69	&	y	&	eb\\
147975720	&	2.8502	&	4.88e+04	&	1.37e+04	&	9.88e+01	&	3464	&	4.73	&	1	&	27.351	&	99.59	&	-37.24	&	y	&	eb\\
33912852	&	1.7049	&	4.78e+04	&	6.74e+04	&	3.69e+01	&	6541	&	4.06	&	2	&	1.9524	&	356.14	&	-26.47	&	y	&	eb\\
35349987	&	10.738	&	4.75e+04	&	1.33e+03	&	1.92e+02	&	3721	&	4.57	&	1	&	19.456	&	224.78	&	-55.66	&	y	&	eb\\
101845679	&	1.2811	&	4.72e+04	&	3.69e+04	&	2.44e+01	&	6252	&	4.06	&	2	&	1.5564	&	304.00	&	-50.13	&	y	&	eb\\
178171080	&	3.2501	&	4.71e+04	&	2.40e+03	&	1.00e+02	&	---	    &	--- 	&	1	&	2.4862	&	43.26	&	-31.40	&	n	&	eb\\
49558810	&	3.3543	&	4.58e+04	&	4.31e+03	&	8.50e+01	&	5209	&	4.34	&	1	&	7.8446	&	36.53	&	-36.37	&	y	&	eb\\
290476605	&	5.7452	&	4.52e+04	&	1.24e+04	&	1.60e+02	&	6687	&	3.96	&	2	&	2.7871	&	139.16	&	0.72	&	y	&	eb\\
140659978	&	9.4407	&	4.46e+04	&	2.25e+03	&	1.41e+02	&	5221	&	4.20	&	2	&	29.384	&	72.45	&	-72.46	&	y	&	eb\\
278784173	&	0.7473	&	4.45e+04	&	2.43e+04	&	3.31e+01	&	---	    &	---	    &	1	&	6.3155	&	335.73	&	-37.52	&	y	&	eb\\
250387838	&	6.8863	&	4.40e+04	&	1.08e+04	&	1.09e+02	&	5871	&	3.80	&	2	&	3.7178	&	30.26	&	-2.49	&	y	&	eb\\
371839073	&	3.7963	&	4.38e+04	&	2.74e+03	&	9.26e+01	&	6034	&	4.22	&	1	&	10.071	&	306.00	&	-71.08	&	y	&	eb\\
269852699	&	5.1365	&	4.38e+04	&	2.12e+03	&	7.51e+01	&	6137	&	4.13	&	2	&	11.706	&	69.88	&	-79.72	&	y	&	eb\\
220397947	&	1.7757	&	4.30e+04	&	3.94e+03	&	5.98e+01	&	6257	&	4.01	&	1	&	7.2456	&	69.01	&	-58.07	&	y	&	eb\\
340100436	&	1.4689	&	4.30e+04	&	1.17e+04	&	5.98e+01	&	5297	&	4.43	&	1	&	3.9916	&	237.69	&	-63.97	&	y	&	eb\\
379783522	&	2.4499	&	4.23e+04	&	1.91e+04	&	4.14e+01	&	---	    &	--- 	&	2	&	1.6508	&	181.59	&	-65.70	&	n	&	eb\\
255700967	&	3.4253	&	4.23e+04	&	2.56e+03	&	1.00e+02	&	6896	&	3.94	&	1	&	4.4857	&	98.90	&	-52.33	&	y	&	eb\\
201747686	&	1.3457	&	4.20e+04	&	1.60e+04	&	3.02e+01	&	6233	&	---	    &	2	&	2.2124	&	303.47	&	-57.99	&	n	&	eb\\
273792220	&	1.9879	&	4.19e+04	&	2.99e+05	&	3.61e+01	&	5892	&	4.13	&	2	&	6.4483	&	30.55	&	-62.59	&	n	&	eb\\
158150633	&	2.0783	&	4.13e+04	&	3.24e+03	&	5.82e+01	&	4525	&	---	    &	2	&	11.701	&	162.13	&	-38.90	&	n	&	eb\\
250196734	&	7.7037	&	4.11e+04	&	1.94e+05	&	1.41e+02	&	3962	&	4.41	&	2	&	50.505	&	65.21	&	-2.62	&	y	&	eb\\
102691227	&	1.2097	&	4.10e+04	&	6.55e+04	&	4.60e+01	&	6457	&	4.05	&	1	&	7.4433	&	155.57	&	-43.21	&	y	&	eb\\
146522930	&	1.0875	&	4.09e+04	&	8.56e+03	&	2.10e+01	&	5812	&	---	    &	2	&	4.0696	&	76.69	&	-19.96	&	y	&	eb\\
278706358	&	3.4512	&	4.08e+04	&	3.42e+04	&	1.15e+02	&	5619	&	4.11	&	2	&	36.414	&	334.81	&	-40.53	&	y	&	eb\\
55369219	&	1.9793	&	4.06e+04	&	5.22e+03	&	4.54e+01	&	6821	&	3.82	&	1	&	7.4981	&	77.50	&	-61.59	&	y	&	eb\\
348759510	&	3.7162	&	4.04e+04	&	2.43e+04	&	1.41e+02	&	6672	&	4.28	&	1	&	2.9118	&	134.20	&	-17.43	&	y	&	eb\\
253715855	&	1.1317	&	4.02e+04	&	1.88e+04	&	3.82e+01	&	6138	&	4.20	&	2	&	4.3179	&	282.06	&	-37.18	&	n	&	eb\\
158582801	&	1.0801	&	4.02e+04	&	4.98e+05	&	3.78e+01	&	---     &	---	    &	1	&	43.263	&	21.86	&	-49.47	&	y	&	eb\\
219362976	&	7.9768	&	4.01e+04	&	4.65e+03	&	2.11e+02	&	4387	&	4.29	&	1	&	20.637	&	75.11	&	-49.42	&	y	&	eb\\
237944385	&	4.4247	&	4.00e+04	&	1.97e+03	&	1.52e+02	&	6304	&	4.03	&	1	&	1.7937	&	101.22	&	-48.70	&	y	&	pl\\
201293780	&	0.9377	&	4.00e+04	&	3.04e+04	&	2.45e+01	&	5962	&	4.29	&	2	&	1.7623	&	3.23	&	-57.35	&	y	&	eb\\
436158814	&	7.2141	&	3.99e+04	&	1.85e+03	&	1.27e+02	&	3510	&	4.64	&	2	&	19.849	&	83.94	&	9.76	&	y	&	eb\\
176591772	&	7.7591	&	3.94e+04	&	1.68e+03	&	1.03e+02	&	5626	&	4.19	&	2	&	6.9878	&	87.57	&	-5.77	&	y	&	eb\\
148612685	&	5.0098	&	3.88e+04	&	6.32e+04	&	9.22e+01	&	---	    &	---	    &	2	&	12.547	&	165.34	&	-11.89	&	y	&	eb\\
293081694	&	1.2540	&	3.83e+04	&	1.95e+04	&	3.80e+01	&	6335	&	4.19	&	2	&	3.1344	&	175.04	&	-27.26	&	n	&	eb\\
92594505	&	11.435	&	3.79e+04	&	1.38e+03	&	1.72e+02	&	6302	&	3.75	&	1	&	7.8543	&	315.46	&	-33.48	&	y	&	eb\\
231293332	&	1.7879	&	3.78e+04	&	5.31e+04	&	6.50e+01	&	6140	&	4.22	&	1	&	15.375	&	44.71	&	-49.17	&	y	&	eb\\
88380001	&	4.4343	&	3.74e+04	&	2.42e+03	&	1.15e+02	&	6458	&	4.10	&	1	&	1.7522	&	44.79	&	-25.23	&	y	&	pl\\
231809798	&	0.9057	&	3.63e+04	&	9.96e+04	&	1.99e+01	&	3520	&	4.74	&	2	&	18.734	&	85.53	&	-52.11	&	y	&	eb\\
281494100	&	2.3076	&	3.63e+04	&	2.38e+05	&	6.51e+01	&	6427	&	4.17	&	2	&	43.686	&	112.19	&	-56.40	&	y	&	eb\\
253657883	&	0.7908	&	3.62e+04	&	9.50e+03	&	2.08e+01	&	4316	&	4.51	&	2	&	8.2538	&	281.47	&	-39.41	&	n	&	eb\\
140659980	&	9.4392	&	3.57e+04	&	1.88e+03	&	1.63e+02	&	5642	&	3.84	&	2	&	5.3466	&	72.47	&	-72.46	&	n	&	eb\\
387544749	&	2.8185	&	3.55e+04	&	3.55e+04	&	6.12e+01	&	3711	&	4.59	&	2	&	9.5871	&	44.25	&	10.31	&	n	&	eb\\
143022688	&	1.5663	&	3.55e+04	&	2.12e+05	&	5.30e+01	&	4086	&	---	    &	2	&	12.777	&	54.26	&	-30.07	&	y	&	eb\\
117739806	&	5.6174	&	3.55e+04	&	8.52e+02	&	1.14e+02	&	5981	&	4.39	&	2	&	20.181	&	96.20	&	-31.86	&	y	&	eb\\
278861826	&	0.5337	&	3.47e+04	&	2.79e+04	&	1.58e+01	&	6369	&	4.03	&	1	&	3.7808	&	100.65	&	-56.54	&	y	&	eb\\
52641430	&	1.6715	&	3.47e+04	&	9.45e+03	&	2.90e+01	&	---	    &	---	    &	2   &	1.6214	&	101.11	&	-33.98	&	n	&	eb\\
39818458	&	3.8154	&	3.44e+04	&	1.85e+04	&	1.25e+02	&	12561	&	--- 	&	1	&	5.3337	&	90.19	&	-3.89	&	y	&	eb\\
211116547	&	1.2768	&	3.44e+04	&	1.43e+04	&	2.92e+01	&	---	    &	4.74	&	2	&	36.378	&	280.43	&	-31.67	&	y	&	eb\\
333642126	&	4.827	&	3.41e+04	&	1.45e+03	&	9.64e+01	&	5051	&	4.30	&	1	&	11.344	&	56.36	&	-4.12	&	y	&	eb\\
48658291	&	1.5878	&	3.40e+04	&	2.00e+04	&	6.16e+01	&	6322	&	3.74	&	1	&	8.8601	&	96.92	&	-25.83	&	y	&	eb\\
198076334	&	0.3913	&	3.39e+04	&	2.74e+05	&	1.53e+01	&	5854	&	3.61	&	1	&	2.4643	&	63.65	&	-59.08	&	y	&	eb\\
3921749	    &	5.6108	&	3.39e+04	&	5.66e+04	&	1.34e+02	&	4007	&	4.38    &	1	&	33.918	&	11.83	&	-6.44	&	y	&	eb\\
253749178	&	1.8377	&	3.38e+04	&	4.19e+03	&	5.09e+01	&	5958	&	---	    &	2	&	10.192	&	282.34	&	-36.76	&	n	&	eb\\
310556491	&	1.5554	&	3.28e+04	&	6.39e+04	&	4.03e+01	&	6902	&	4.15	&	2	&	5.2244	&	185.28	&	-65.83	&	n	&	eb\\
77437462	&	0.6079	&	3.27e+04	&	5.41e+03	&	3.11e+01	&	3588	&	4.77	&	1	&	1.8442	&	73.13	&	-36.53	&	n	&	pl\\
294059372	&	0.6666	&	3.26e+04	&	3.27e+04	&	1.95e+01	&	6393	&	---     &	2	&	7.1110	&	263.93	&	-71.40	&	n	&	eb\\
300972613	&	0.3880	&	3.25e+04	&	1.34e+05	&	1.30e+01	&	---	    &	---	    &	2	&	6.5351	&	10.77	&	-47.30	&	y	&	eb\\
357911163	&	4.7821	&	3.22e+04	&	2.19e+04	&	8.52e+01	&	6682	&	4.42	&	2	&	15.130	&	192.89	&	-79.02	&	y	&	eb\\
25188036	&	3.3930	&	3.18e+04	&	1.21e+03	&	4.41e+01	&	2868	&	5.05	&	2	&	1009.9	&	98.02	&	-7.31	&	y	&	eb\\
268587594	&	9.1404	&	3.18e+04	&	2.40e+03	&	2.00e+02	&	5887	&	4.19	&	3	&	3.8126	&	15.73	&	-23.88	&	n	&	misc\\
278826996	&	9.5834	&	3.16e+04	&	1.39e+03	&	1.77e+02	&	5800	&	4.03	&	2	&	35.039	&	100.33	&	-54.02	&	y	&	eb\\
52641431	&	0.8358	&	3.16e+04	&	7.13e+03	&	2.25e+01	&	---	    &	---	    &	2	&	1.5964	&	101.11	&	-33.98	&	n	&	eb\\
434664688	&	8.4292	&	3.09e+04	&	1.90e+03	&	1.54e+02	&	---	    &	---	    &	2	&	18.548	&	191.23	&	-65.31	&	y	&	eb\\
142148228	&	0.7805	&	3.06e+04	&	9.43e+03	&	2.27e+01	&	6775	&	4.08	&	1	&	0.6429	&	100.36	&	-74.67	&	n	&	pl\\
150357064	&	1.4894	&	3.03e+04	&	3.55e+03	&	4.98e+01	&	5722	&	4.13	&	1	&	5.8414	&	96.10	&	-64.49	&	n	&	eb\\
243475355	&	1.0957	&	3.03e+04	&	1.26e+04	&	3.07e+01	&	13560	&	---	    &	2	&	1.2504	&	205.91	&	-42.06	&	y	&	eb\\
146439804	&	11.014	&	3.03e+04	&	6.27e+02	&	1.74e+02	&	3801	&	4.32	&	1	&	10.465	&	75.73	&	-22.22	&	y	&	eb\\
262696097	&	0.9111	&	3.03e+04	&	7.64e+03	&	2.69e+01	&	6782	&	3.79	&	1	&	3.1139	&	107.78	&	7.45	&	n	&	eb\\
424992973	&	6.9254	&	3.02e+04	&	6.41e+03	&	1.83e+02	&	5171	&	---	    &	2	&	35.136	&	81.80	&	-18.57	&	y	&	eb\\
38937499	&	2.0276	&	3.00e+04	&	2.65e+03	&	5.70e+01	&	5199	&	4.32	&	2	&	0.3917	&	10.30	&	-68.45	&	y	&	eb\\
79548760	&	7.9728	&	2.98e+04	&	9.98e+02	&	1.43e+02	&	6452	&	3.66	&	1	&	4.0596	&	51.07	&	-45.09	&	y	&	eb\\
143171265	&	1.1410	&	2.97e+04	&	1.22e+04	&	3.72e+01	&	7333	&	4.31	&	2	&	1.7917	&	89.99	&	-34.45	&	n	&	eb\\
294092960	&	0.9022	&	2.97e+04	&	1.54e+04	&	2.23e+01	&	5766	&	4.57	&	2	&	5.0331	&	107.35	&	-57.50	&	n	&	eb\\
146688260	&	0.5030	&	2.94e+04	&	3.01e+04	&	1.53e+01	&	5551	&	4.28	&	2	&	1.3734	&	43.22	&	-43.52	&	n	&	eb\\
58466459	&	3.8717	&	2.93e+04	&	6.87e+02	&	6.60e+01	&	6398	&	4.09	&	2	&	7.2758	&	201.66	&	-27.14	&	y	&	eb\\
148544875	&	1.7441	&	2.92e+04	&	8.11e+03	&	6.30e+01	&	3159	&	4.88	&	1	&	24.273	&	164.12	&	-9.83	&	y	&	eb\\
332488398	&	1.0416	&	2.92e+04	&	8.54e+04	&	2.70e+01	&	5458	&	4.29	&	1	&	4.7379	&	350.56	&	-29.70	&	y	&	eb\\
281728376	&	11.177	&	2.91e+04	&	8.26e+02	&	1.73e+02	&	6656	&	4.00	&	1	&	2.6819	&	11.65	&	-54.10	&	y	&	eb\\
317411334	&	1.4485	&	2.89e+04	&	1.48e+04	&	4.88e+01	&	6036	&	4.15	&	1	&	8.9622	&	87.71	&	-12.15	&	y	&	eb\\
53997651	&	1.4587	&	2.86e+04	&	4.79e+04	&	6.27e+01	&	3418	&	4.71	&	1	&	73.489	&	159.57	&	-33.26	&	y	&	eb\\
441422220	&	3.6995	&	2.81e+04	&	6.93e+03	&	1.13e+02	&	6348	&	4.26	&	2	&	28.628	&	311.36	&	-34.64	&	y	&	eb\\
284196801	&	11.158	&	2.80e+04	&	1.03e+03	&	1.72e+02	&	6098	&	3.77	&	2	&	3.6876	&	105.54	&	-71.12	&	y	&	eb\\
90716944	&	8.0234	&	2.79e+04	&	6.12e+03	&	2.12e+02	&	4900	&	4.55	&	2	&	15.891	&	277.08	&	-45.49	&	y	&	eb\\
386262459	&	2.6672	&	2.79e+04	&	2.00e+03	&	8.05e+01	&	9557	&	4.15	&	2	&	1.9070	&	127.40	&	-16.00	&	n	&	eb\\
91246422	&	9.4234	&	2.76e+04	&	2.73e+03	&	1.61e+02	&	12417	&	---	    &	1	&	4.2674	&	277.93	&	-45.91	&	y	&	eb\\
278826516	&	7.0430	&	2.75e+04	&	5.33e+02	&	1.24e+02	&	6248	&	4.10	&	1	&	10.377	&	100.47	&	-54.81	&	y	&	eb\\
91636891	&	13.450	&	2.74e+04	&	9.44e+02	&	1.61e+02	&	6461	&	3.91	&	1	&	20.549	&	39.40	&	-36.97	&	y	&	eb\\
153742549	&	6.0700	&	2.74e+04	&	2.10e+03	&	8.23e+01	&	5726	&	3.63	&	2	&	11.324	&	66.52	&	-44.59	&	y	&	eb\\
178723395	&	1.7616	&	2.70e+04	&	2.30e+03	&	5.09e+01	&	6756	&	4.26	&	1	&	4.0281	&	166.14	&	-43.24	&	y	&	eb\\
279949020	&	6.5493	&	2.65e+04	&	1.04e+03	&	1.37e+02	&	5860	&	3.77	&	2	&	14.611	&	40.39	&	-59.98	&	y	&	eb\\
207080350	&	9.1069	&	2.65e+04	&	3.52e+05	&	1.93e+02	&	6978	&	3.98	&	2	&	33.987	&	320.91	&	-39.77	&	y	&	eb\\
27654882	&	1.1156	&	2.65e+04	&	2.99e+03	&	3.74e+01	&	6447	&	3.88	&	1	&	3.4932	&	123.70	&	12.06	&	y	&	eb\\
270204401	&	2.2383	&	2.64e+04	&	3.58e+03	&	4.77e+01	&	9428	&	4.19	&	2	&	17.609	&	332.71	&	-71.97	&	y	&	eb\\
350297040	&	10.874	&	2.63e+04	&	4.47e+02	&	1.27e+02	&	6294	&	4.45	&	2	&	19.383	&	83.71	&	-55.93	&	y	&	eb\\
12158853	&	0.6311	&	2.61e+04	&	2.55e+05	&	2.17e+01	&	3465	&	4.69	&	2	&	18.312	&	91.55	&	-3.61	&	y	&	eb\\
218152299	&	1.2597	&	2.59e+04	&	1.63e+04	&	3.65e+01	&	6225	&	3.83	&	1	&	7.6603	&	261.91	&	-45.55	&	y	&	eb\\
2372579	    &	9.9215	&	2.58e+04	&	2.30e+03	&	2.16e+02	&	5417	&	4.22	&	1	&	16.004	&	288.91	&	-28.68	&	y	&	eb\\
32449963	&	6.5199	&	2.58e+04	&	5.22e+04	&	1.91e+02	&	5771	&	4.27	&	2	&	31.714	&	189.93	&	-11.64	&	y	&	eb\\
70717462	&	2.9482	&	2.52e+04	&	2.18e+03	&	8.70e+01	&	---     &	--- 	&	1	&	2.2676	&	151.63	&	-35.18	&	n	&	eb\\
426541180	&	4.6975	&	2.50e+04	&	1.15e+03	&	9.95e+01	&	5715	&	4.32	&	1	&	4.7776	&	244.93	&	-72.21	&	n	&	eb\\
358108063	&	2.9858	&	2.49e+04	&	1.06e+03	&	7.24e+01	&	5723	&	---	    &	2	&	8.8750	&	51.24	&	-75.93	&	y	&	eb\\
394585218	&	3.9645	&	2.48e+04	&	1.46e+03	&	9.38e+01	&	6032	&	3.99	&	2	&	8.3107	&	341.02	&	-65.76	&	y	&	eb\\
178996712	&	2.9828	&	2.46e+04	&	1.46e+04	&	4.16e+01	&	7188	&	4.10	&	2	&	3.6709	&	70.06	&	-27.73	&	n	&	eb\\
220359766	&	1.7599	&	2.45e+04	&	2.14e+03	&	4.65e+01	&	--- 	&	--- 	&	1	&	2.2444	&	155.05	&	-52.19	&	y	&	eb\\
341309569	&	7.3475	&	2.45e+04	&	6.00e+02	&	1.24e+02	&	4966	&	---	    &	1	&	4.7958	&	315.92	&	-64.13	&	n	&	eb\\
139188326	&	0.5883	&	2.42e+04	&	8.99e+04	&	1.08e+01	&	6506	&	3.67	&	2	&	0.7701	&	345.01	&	-46.49	&	n	&	eb\\
270166263	&	7.6589	&	2.41e+04	&	2.03e+03	&	1.63e+02	&	4933	&	4.13	&	2	&	11.119	&	330.03	&	-27.70	&	y	&	eb\\
317191077	&	2.1910	&	2.41e+04	&	4.85e+03	&	4.17e+01	&	6364	&	---	    &	2	&	4.7310	&	87.14	&	-13.05	&	n	&	eb\\
409900613	&	1.1032	&	2.40e+04	&	9.07e+03	&	4.67e+01	&	3194	&	4.83	&	1	&	9.5539	&	144.89	&	-17.59	&	y	&	eb\\
118200081	&	0.9529	&	2.39e+04	&	1.01e+04	&	1.79e+01	&	5366	&	---	    &	2	&	3.4996	&	272.18	&	-56.78	&	n	&	eb\\
115275539	&	1.7630	&	2.37e+04	&	3.17e+03	&	5.11e+01	&	4913	&	--- 	&	1	&	2.3541	&	25.95	&	-38.72	&	y	&	eb\\
46313508	&	1.0326	&	2.33e+04	&	6.23e+03	&	2.50e+01	&	6445	&	3.60	&	2	&	2.3889	&	83.10	&	-14.97	&	n	&	eb\\
124350360	&	8.5580	&	2.33e+04	&	2.38e+04	&	2.33e+02	&	3471	&	4.62	&	1	&	14.902	&	227.66	&	-52.80	&	y	&	eb\\
169316288	&	0.5589	&	2.33e+04	&	4.26e+04	&	1.42e+01	&	6290	&	---	    &	2	&	5.4193	&	292.50	&	-27.95	&	n	&	eb\\
222945452	&	0.7483	&	2.31e+04	&	1.41e+04	&	2.12e+01	&	---	    &	--- 	&	2	&	0.9521	&	245.89	&	-44.11	&	n	&	eb\\
228760807	&	0.7779	&	2.30e+04	&	9.22e+04	&	2.03e+01	&	5320	&	--- 	&	2	&	3.9835	&	194.10	&	-27.39	&	y	&	eb\\
396655059	&	7.9781	&	2.30e+04	&	1.17e+03	&	1.00e+02	&	7909	&	4.16	&	2	&	12.262	&	214.77	&	-77.09	&	y	&	eb\\
440788366	&	0.7742	&	2.28e+04	&	6.48e+03	&	2.05e+01	&	9079	&	---	    &	1	&	2.6778	&	109.09	&	12.45	&	n	&	eb\\
18914916	&	11.287	&	2.27e+04	&	5.66e+02	&	1.51e+02	&	5486	&	4.57	&	2	&	15.729	&	118.92	&	13.80	&	y	&	eb\\
180943676	&	8.418	&	2.26e+04	&	8.62e+02	&	1.52e+02	&	6027	&	4.39	&	1	&	8.2751	&	88.72	&	-46.42	&	n	&	eb\\
294197236	&	0.9672	&	2.24e+04	&	4.36e+03	&	3.07e+01	&	6372	&	3.60	&	1	&	5.6588	&	264.81	&	-68.78	&	n	&	eb\\
312111256	&	1.1517	&	2.19e+04	&	1.01e+04	&	2.71e+01	&	7593	&	4.38	&	2	&	3.6697	&	140.66	&	6.23	&	n	&	eb\\
423524230	&	2.5815	&	2.18e+04	&	1.32e+03	&	6.25e+01	&	6564	&	4.33	&	2	&	2.7988	&	169.18	&	-19.89	&	n	&	eb\\
30382918	&	0.5719	&	2.17e+04	&	2.86e+05	&	1.04e+01	&	5854	&	4.22	&	2	&	3.3986	&	198.43	&	-41.27	&	y	&	eb\\
48063034	&	0.5863	&	2.16e+04	&	5.19e+04	&	1.52e+01	&	5304	&	4.20	&	2	&	6.5276	&	147.48	&	-34.11	&	n	&	eb\\
67267845	&	1.0126	&	2.16e+04	&	4.29e+03	&	3.57e+01	&	6817	&	3.64	&	1	&	11.205	&	240.95	&	-29.97	&	n	&	eb\\
35101462	&	4.1726	&	2.16e+04	&	1.76e+03	&	7.34e+01	&	6114	&	4.21	&	2	&	5.3038	&	35.41	&	-14.78	&	y	&	eb\\
363933313	&	4.6865	&	2.13e+04	&	9.98e+02	&	9.37e+01	&	5428	&	--- 	&	1	&	11.931	&	251.32	&	-71.05	&	y	&	eb\\
270411008	&	8.6657	&	2.11e+04	&	2.14e+04	&	2.31e+02	&	5492	&	4.24	&	1	&	41.883	&	349.40	&	-29.17	&	y	&	eb\\
301247295	&	8.5054	&	2.10e+04	&	1.47e+03	&	9.64e+01	&	5933	&	---	    &	2	&	15.232	&	255.08	&	-69.56	&	y	&	eb\\
244163413	&	1.7879	&	2.09e+04	&	8.52e+04	&	6.67e+01	&	6098	&	4.29	&	2	&	14.971	&	5.16	&	-5.14	&	y	&	eb\\
468958384	&	0.7675	&	2.07e+04	&	4.07e+03	&	2.16e+01	&	3478	&	4.81	&	1	&	35.711	&	107.70	&	7.90	&	y	&	eb\\
146761368	&	0.9176	&	2.05e+04	&	3.53e+03	&	2.83e+01	&	6597	&	3.58	&	1	&	6.5819	&	46.17	&	-44.66	&	n	&	eb\\
191926695	&	1.9352	&	2.02e+04	&	9.92e+03	&	3.56e+01	&	---	    &	--- 	&	2	&	2.2496	&	136.52	&	-38.88	&	y	&	eb\\
303427297	&	4.2976	&	2.02e+04	&	1.29e+03	&	5.27e+01	&	5273	&	--- 	&	2	&	1.7764	&	164.18	&	-57.75	&	n	&	eb\\
260659986	&	3.5699	&	2.01e+04	&	9.93e+02	&	9.34e+01	&	6107	&	3.90	&	1	&	16.691	&	98.48	&	-59.96	&	y	&	eb\\
117879940	&	1.1311	&	2.00e+04	&	3.29e+03	&	2.72e+01	&	5559	&	3.92	&	2	&	3.8599	&	70.72	&	-11.87	&	n	&	eb\\
380645083	&	4.3894	&	1.99e+04	&	1.95e+03	&	8.61e+01	&	5927	&	3.58	&	2	&	0.9681	&	285.20	&	-34.88	&	n	&	eb\\
259864042	&	0.4607	&	1.99e+04	&	2.81e+03	&	3.90e+01	&	31000	&	5.59	&	1	&	14.676	&	40.32	&	-68.92	&	n	&	eb\\
234494090	&	0.9726	&	1.96e+04	&	3.10e+03	&	2.98e+01	&	---	    &	---	    &	1	&	19.558	&	9.81	&	-63.20	&	y	&	eb\\
319467578	&	2.3652	&	1.93e+04	&	1.48e+03	&	4.72e+01	&	6157	&	4.22	&	2	&	2.4735	&	100.22	&	-52.69	&	y	&	eb\\
369966331	&	0.9128	&	1.92e+04	&	8.60e+03	&	1.77e+01	&	4686	&	--- 	&	2	&	0.9706	&	19.73	&	-65.18	&	y	&	eb\\
291320884	&	0.3717	&	1.88e+04	&	3.30e+05	&	8.40e+00	&	3507	&	4.73	&	2	&	33.227	&	260.95	&	-57.41	&	n	&	eb\\
410654298	&	8.2252	&	1.79e+04	&	1.91e+04	&	2.18e+02	&	5234	&	--- 	&	1	&	18.858	&	177.00	&	-66.11	&	y	&	eb\\
66509654	&	2.6446	&	1.76e+04	&	3.45e+03	&	9.16e+01	&	3253	&	4.80	&	1	&	35.119	&	109.42	&	-23.99	&	y	&	eb\\
22937402	&	1.3839	&	1.76e+04	&	1.59e+04	&	2.29e+01	&	---	    &	--- 	&	2	&	0.6199	&	109.13	&	-38.31	&	n	&	eb\\
94257578	&	1.1921	&	1.74e+04	&	2.43e+03	&	5.15e+01	&	5722	&	---	    &	2	&	3.2446	&	241.41	&	-30.64	&	n	&	eb\\
263925187	&	0.3330	&	1.73e+04	&	2.71e+04	&	1.21e+01	&	3944	&	4.59	&	1	&	21.136	&	269.99	&	-30.07	&	n	&	eb\\
9053429	    &	12.962	&	1.72e+04	&	6.14e+02	&	1.30e+02	&	31000	&	5.59	&	1	&	33.615	&	356.09	&	-10.19	&	y	&	eb\\
197886566	&	0.8721	&	1.71e+04	&	1.32e+04	&	1.45e+01	&	6019	&	4.04	&	1	&	1.1038	&	57.31	&	-58.44	&	n	&	pl\\
11756637	&	4.3061	&	1.70e+04	&	2.38e+04	&	9.04e+01	&	3436	&	4.60	&	2	&	53.312	&	90.61	&	-1.31	&	y	&	eb\\
78450017	&	0.9438	&	1.70e+04	&	2.19e+03	&	2.75e+01	&	3524	&	4.60	&	1	&	11.030	&	102.58	&	-21.46	&	y	&	eb\\
259543079	&	9.0208	&	1.69e+04	&	5.39e+02	&	1.24e+02	&	5241	&	4.31	&	1	&	17.720	&	71.63	&	-53.12	&	y	&	eb\\
365570620	&	0.8584	&	1.69e+04	&	3.59e+03	&	2.69e+01	&	6505	&	3.73	&	1	&	4.8005	&	81.29	&	7.95	&	n	&	eb\\
142177867	&	1.4353	&	1.65e+04	&	7.92e+04	&	3.44e+01	&	5456	&	4.33	&	2	&	3.8672	&	37.41	&	-47.22	&	y	&	eb\\
61724376	&	0.9538	&	1.64e+04	&	3.60e+03	&	2.46e+01	&	5780	&	3.82	&	1	&	5.1693	&	288.95	&	-44.50	&	n	&	eb\\
220015845	&	54.515	&	1.64e+04	&	2.43e+03	&	2.10e+02	&	5950	&	4.12	&	1	&	33.659	&	32.36	&	-49.55	&	n	&	eb\\
173038281	&	4.309	&	1.64e+04	&	1.00e+03	&	8.94e+01	&	6412	&	4.22	&	1	&	1.8536	&	102.27	&	-27.34	&	y	&	pl\\
60479151	&	0.4189	&	1.63e+04	&	1.03e+04	&	1.46e+01	&	6210	&	3.99	&	1	&	1.6225	&	96.31	&	-21.78	&	y	&	pl\\
164681172	&	4.9064	&	1.61e+04	&	4.12e+04	&	1.09e+02	&	5694	&	4.23	&	2	&	31.586	&	24.56	&	-21.07	&	y	&	eb\\
91961   	&	5.2823	&	1.56e+04	&	9.56e+02	&	8.55e+01	&	---	    &	4.83	&	2	&	18.651	&	221.69	&	-23.91	&	y	&	eb\\
31690845	&	1.1991	&	1.54e+04	&	6.68e+03	&	1.77e+01	&	6652	&	3.43	&	2	&	2.2371	&	289.12	&	-31.40	&	n	&	eb\\
48063032	&	0.5863	&	1.54e+04	&	2.84e+04	&	1.51e+01	&	5824	&	4.39	&	2	&	5.3075	&	147.48	&	-34.11	&	n	&	eb\\
162585282	&	0.5122	&	1.53e+04	&	1.02e+05	&	8.79e+00	&	3311	&	4.75	&	2	&	31.019	&	170.31	&	-47.58	&	n	&	eb\\
4723156 	&	9.4896	&	1.51e+04	&	4.85e+04	&	1.72e+02	&	3057	&	4.96	&	2	&	41.058	&	39.44	&	-7.09	&	y	&	eb\\
323295967	&	0.4245	&	1.50e+04	&	1.59e+04	&	1.15e+01	&	5581	&	--- 	&	2	&	1.8769	&	52.64	&	-39.24	&	n	&	eb\\
219328784	&	1.9687	&	1.47e+04	&	1.55e+03	&	4.42e+01	&	4539	&	4.30	&	1	&	1.0865	&	333.50	&	-54.11	&	n	&	eb\\
18915591	&	3.4933	&	1.45e+04	&	8.48e+02	&	8.28e+01	&	7432	&	3.41	&	1	&	21.963	&	118.91	&	12.66	&	y	&	eb\\
12631605	&	0.7086	&	1.44e+04	&	1.28e+05	&	1.66e+01	&	5514	&	4.44	&	2	&	20.219	&	53.11	&	-6.97	&	y	&	eb\\
147314529	&	0.4628	&	1.40e+04	&	1.08e+04	&	1.67e+01	&	11310	&	3.70	&	1	&	1.3431	&	104.69	&	-17.63	&	n	&	pl\\
340633943	&	0.3757	&	1.40e+04	&	4.79e+04	&	9.30e+00	&	6031	&	4.41	&	1	&	6.0757	&	116.49	&	-56.54	&	n	&	eb\\
412345587	&	0.5268	&	1.38e+04	&	3.06e+04	&	1.72e+01	&	3516	&	4.82	&	1	&	121.13	&	110.53	&	-18.64	&	n	&	eb\\
114819301	&	4.8876	&	1.37e+04	&	8.06e+02	&	7.87e+01	&	3786	&	4.32	&	1	&	6.2994	&	2.30	&	-18.91	&	y	&	eb\\
401926767	&	6.8009	&	1.36e+04	&	1.26e+05	&	1.20e+02	&	4342	&	4.25	&	2	&	48.278	&	57.35	&	-1.24	&	y	&	eb\\
52924466	&	10.612	&	1.34e+04	&	3.99e+02	&	1.31e+02	&	5305	&	4.43	&	1	&	3.5125	&	102.38	&	-33.70	&	y	&	eb\\
350298314	&	13.525	&	1.33e+04	&	3.58e+02	&	1.22e+02	&	5468	&	4.59	&	1	&	23.465	&	83.74	&	-59.33	&	y	&	eb\\
183537458	&	0.2278	&	1.32e+04	&	2.00e+05	&	7.49e+00	&	6283	&	4.42	&	1	&	1.8545	&	357.79	&	-39.86	&	y	&	usp\\
159897979	&	16.785	&	1.29e+04	&	1.29e+03	&	1.12e+02	&	6687	&	3.74	&	2	&	20.779	&	48.84	&	-38.99	&	y	&	eb\\
343757299	&	5.8194	&	1.28e+04	&	9.17e+02	&	7.35e+01	&	9787	&	4.00	&	1	&	5.1452	&	273.21	&	-74.28	&	y	&	eb\\
219201178	&	2.4170	&	1.28e+04	&	7.41e+02	&	4.50e+01	&	6760	&	4.03	&	2	&	3.6424	&	92.72	&	-52.89	&	n	&	eb\\
148611095	&	4.8374	&	1.27e+04	&	1.01e+04	&	5.47e+01	&	3452	&	4.78	&	2	&	13.377	&	165.21	&	-7.26	&	y	&	eb\\
143234741	&	1.9178	&	1.27e+04	&	7.46e+02	&	3.75e+01	&	5855	&	4.17	&	1	&	3.5656	&	90.62	&	-32.38	&	n	&	eb\\
167812449	&	4.2356	&	1.25e+04	&	4.60e+02	&	7.52e+01	&	5799	&	4.28	&	2	&	1.7238	&	104.78	&	-61.84	&	n	&	eb\\
269996393	&	10.381	&	1.25e+04	&	4.94e+02	&	1.28e+02	&	5711	&	4.51	&	1	&	0.5600	&	311.46	&	-27.60	&	n	&	misc\\
358022169	&	0.3032	&	1.24e+04	&	3.48e+04	&	8.15e+00	&	5740	&	4.53	&	2	&	0.5261	&	193.65	&	-80.07	&	y	&	eb\\
159803421	&	0.2761	&	1.24e+04	&	1.32e+04	&	1.23e+01	&	5708	&	4.37	&	1	&	0.2218	&	261.64	&	-32.96	&	y	&	usp\\
120566395	&	1.7957	&	1.23e+04	&	1.72e+03	&	2.67e+01	&	6001	&	4.04	&	1	&	1.3061	&	3.84	&	-39.52	&	n	&	pl\\
124712813	&	0.8256	&	1.23e+04	&	2.30e+03	&	2.07e+01	&	3592	&	4.68	&	2	&	7.0125	&	104.76	&	-11.72	&	n	&	eb\\
326356701	&	1.1122	&	1.20e+04	&	4.50e+03	&	2.18e+01	&	4674	&	--- 	&	2	&	1.5596	&	325.59	&	-21.76	&	n	&	eb\\
355357949	&	1.4604	&	1.17e+04	&	1.20e+03	&	3.51e+01	&	6714	&	4.17	&	1	&	3.5642	&	109.33	&	-49.41	&	n	&	eb\\
47175622	&	1.5235	&	1.16e+04	&	2.62e+03	&	2.84e+01	&	6698	&	---	    &	2	&	1.5101	&	145.35	&	-7.65	&	y	&	eb\\
293480903	&	7.5701	&	1.11e+04	&	5.00e+02	&	7.90e+01	&	6180	&	---	    &	2	&	3.4731	&	49.97	&	-82.14	&	n	&	eb\\
63037741	&	12.295	&	1.09e+04	&	3.03e+02	&	1.52e+02	&	3341	&	4.84	&	1	&	6.5 	&	336.86	&	-35.01	&	y	&	misc\\
266769521	&	0.5362	&	1.07e+04	&	1.72e+06	&	7.77e+00	&	6659	&	4.06	&	2	&	36.167	&	29.75	&	-22.91	&	n	&	eb\\
265551788	&	10.943	&	1.04e+04	&	3.55e+02	&	1.19e+02	&	5258	&	---	    &	1	&	1.4161	&	94.75	&	5.93	&	y	&	pl\\
21184505	&	5.1777	&	1.02e+04	&	6.55e+02	&	1.01e+02	&	6146	&	4.23	&	3	&	0.7600	&	128.85	&	12.01	&	n	&	eb\\
443494351	&	0.2578	&	1.02e+04	&	3.05e+04	&	7.87e+00	&	7620	&	3.33	&	1	&	1.8084	&	146.78	&	-54.20	&	n	&	usp\\
425864141	&	0.1091	&	1.01e+04	&	1.60e+05	&	6.38e+00	&	3039	&	4.92	&	1	&	15.052	&	352.08	&	-45.91	&	n	&	usp\\
76698707	&	0.2639	&	1.01e+04	&	1.44e+04	&	9.62e+00	&	5845	&	3.61	&	1	&	1.1410	&	229.05	&	-32.49	&	y	&	usp\\
300560295	&	10.292	&	1.00e+04	&	2.55e+02	&	1.00e+02	&	5885	&	4.25	&	1	&	2.3754	&	112.91	&	-68.13	&	n	&	eb\\
274229418	&	9.6186	&	1.00e+04	&	1.16e+04	&	2.37e+02	&	3422	&	4.76	&	1	&	28.221	&	115.47	&	5.04	&	y	&	eb\\
278683641	&	4.2420	&	9.98e+03	&	1.09e+03	&	5.11e+01	&	6095	&	4.03	&	2	&	14.176	&	99.24	&	-58.46	&	y	&	eb\\
349970685	&	1.2570	&	9.95e+03	&	1.15e+03	&	2.03e+01	&	---	    &	--- 	&	2	&	9.6550	&	114.48	&	-60.29	&	y	&	eb\\
75352949	&	0.4720	&	9.84e+03	&	5.46e+04	&	8.10e+00	&	4816	&	--- 	&	2	&	4.3822	&	6.41	&	-17.81	&	n	&	eb\\
150357290	&	0.4926	&	9.81e+03	&	4.92e+03	&	1.36e+01	&	5650	&	4.39	&	1	&	0.6642	&	96.16	&	-64.05	&	n	&	misc\\
224283342	&	0.8873	&	9.74e+03	&	6.83e+03	&	1.16e+01	&	3165	&	4.88	&	3	&	5.1680	&	356.35	&	-40.33	&	n	&	misc\\
56128191	&	3.1118	&	9.72e+03	&	6.14e+02	&	5.59e+01	&	6798	&	4.20	&	2	&	2.9522	&	70.75	&	-6.43	&	y	&	eb\\
133105082	&	0.4464	&	9.41e+03	&	3.23e+04	&	5.79e+00	&	3282	&	4.74	&	2	&	307.51	&	120.46	&	-40.50	&	n	&	eb\\
309025566	&	8.8084	&	9.41e+03	&	5.13e+04	&	2.38e+02	&	6459	&	---	    &	1	&	13.707	&	151.07	&	-28.37	&	y	&	eb\\
384742846	&	6.3568	&	9.34e+03	&	6.43e+02	&	6.10e+01	&	6542	&	4.20	&	2	&	19.510	&	258.72	&	-82.85	&	y	&	eb\\
173077938	&	1.2970	&	9.33e+03	&	3.07e+03	&	1.66e+01	&	7382	&	3.87	&	2	&	0.3219	&	111.05	&	-40.86	&	n	&	eb\\
254039358	&	0.7477	&	9.16e+03	&	2.33e+03	&	1.60e+01	&	6124	&	4.10	&	1	&	0.4260	&	285.25	&	-40.38	&	y	&	pl\\
121048789	&	0.6441	&	9.16e+03	&	4.15e+03	&	1.80e+01	&	5709	&	4.21	&	1	&	0.3975	&	56.19	&	-21.29	&	y	&	pl\\
308184924	&	4.1283	&	8.89e+03	&	7.20e+02	&	4.39e+01	&	6082	&	3.99	&	2	&	26.963	&	130.43	&	-69.78	&	y	&	eb\\
153709888	&	2.1663	&	8.85e+03	&	9.29e+02	&	3.93e+01	&	6720	&	4.22	&	1	&	6.3590	&	65.56	&	-42.07	&	y	&	eb\\
304071827	&	1.5913	&	8.78e+03	&	2.92e+03	&	2.34e+01	&	6645	&	3.97	&	1	&	0.4086	&	258.45	&	-65.68	&	n	&	pl\\
22146154	&	6.0609	&	8.58e+03	&	2.79e+02	&	8.65e+01	&	6874	&	3.97	&	1	&	2.9735	&	148.76	&	-25.13	&	y	&	eb\\
189639080	&	1.5666	&	8.36e+03	&	4.35e+02	&	3.27e+01	&	8840	&	--- 	&	1	&	18.359	&	144.21	&	-28.48	&	y	&	eb\\
32050581	&	1.5564	&	8.30e+03	&	5.09e+02	&	3.36e+01	&	5503	&	4.22	&	1	&	2.4356	&	57.07	&	-69.15	&	y	&	eb\\
98658304	&	5.9320	&	8.29e+03	&	5.72e+02	&	7.09e+01	&	4996	&	---	    &	2	&	15.442	&	8.93	&	-22.74	&	n	&	eb\\
220569051	&	2.7779	&	8.25e+03	&	6.91e+02	&	4.67e+01	&	6025	&	3.73	&	2	&	6.6498	&	46.36	&	-60.36	&	y	&	eb\\
55659311	&	5.595	&	7.96e+03	&	3.12e+02	&	6.63e+01	&	5103	&	4.45	&	2	&	20.880	&	74.11	&	-64.48	&	y	&	eb\\
408512382	&	4.0311	&	7.96e+03	&	7.19e+02	&	6.06e+01	&	7126	&	3.32	&	2	&	5.8634	&	194.65	&	-21.71	&	y	&	eb\\
219421171	&	8.3023	&	7.89e+03	&	3.44e+02	&	9.05e+01	&	6245	&	3.57	&	2	&	16.753	&	79.82	&	-49.64	&	y	&	eb\\
146344020	&	2.4945	&	7.88e+03	&	7.28e+02	&	2.91e+01	&	5986	&	3.94	&	2	&	2.5643	&	74.34	&	-20.96	&	n	&	eb\\
281053777	&	11.019	&	7.86e+03	&	3.55e+04	&	2.24e+02	&	6927	&	---	    &	1	&	126.47	&	98.82	&	4.19	&	y	&	eb\\
409942986	&	0.2796	&	7.81e+03	&	2.22e+05	&	6.20e+00	&	5528	&	4.22	&	1	&	6.2729	&	145.00	&	-17.16	&	n	&	usp\\
115693377	&	4.9269	&	7.67e+03	&	4.57e+02	&	6.05e+01	&	6136	&	4.13	&	2	&	3.7075	&	6.81	&	-41.57	&	n	&	eb\\
460820330	&	5.0743	&	7.65e+03	&	4.49e+02	&	5.51e+01	&	6114	&	4.35	&	2	&	38.855	&	219.39	&	-31.63	&	y	&	eb\\
270648838	&	7.6347	&	7.56e+03	&	3.17e+04	&	1.58e+02	&	6093	&	4.14	&	2	&	50.901	&	136.70	&	5.46	&	y	&	eb\\
140212820	&	0.7777	&	7.55e+03	&	1.34e+03	&	1.58e+01	&	6392	&	3.94	&	2	&	2.1021	&	84.96	&	-30.76	&	y	&	eb\\
270348090	&	2.8190	&	7.55e+03	&	4.59e+02	&	4.96e+01	&	---	    &	--- 	&	1	&	20.399	&	215.48	&	-30.84	&	y	&	eb\\
269342892	&	0.2244	&	7.52e+03	&	1.31e+04	&	6.97e+00	&	---	    &	---	    &	1	&	0.5568	&	119.07	&	-48.24	&	n	&	usp\\
82580956	&	0.9471	&	7.48e+03	&	1.95e+03	&	2.28e+01	&	5793	&	3.93	&	1	&	0.4057	&	169.62	&	-53.71	&	y	&	pl\\
255548311	&	4.9631	&	7.45e+03	&	2.55e+02	&	5.80e+01	&	---	    &	--- 	&	1	&	3.5553	&	95.48	&	-50.92	&	n	&	eb\\
146559012	&	2.6411	&	7.43e+03	&	3.54e+04	&	8.64e+01	&	6216	&	4.39	&	1	&	17.012	&	77.51	&	-23.67	&	y	&	eb\\
92628706	&	9.2030	&	7.42e+03	&	1.08e+05	&	2.30e+02	&	5547	&	4.20	&	1	&	30.235	&	315.74	&	-32.58	&	y	&	eb\\
107012110	&	3.0334	&	7.34e+03	&	5.32e+02	&	5.40e+01	&	5994	&	4.14	&	1	&	1.4840	&	162.04	&	-40.00	&	n	&	pl\\
451211626	&	2.4527	&	7.04e+03	&	6.29e+02	&	3.24e+01	&	3773	&	4.63	&	2	&	41.088	&	210.31	&	-34.96	&	y	&	eb\\
279537689	&	5.7846	&	7.01e+03	&	5.36e+02	&	8.19e+01	&	5958	&	4.26	&	2	&	3.9806	&	59.40	&	-14.16	&	y	&	eb\\
9172417 	&	0.2451	&	6.89e+03	&	3.80e+05	&	5.19e+00	&	3551	&	4.67	&	2	&	6.4761	&	62.05	&	-5.75	&	n	&	usp\\
405391996	&	1.2851	&	6.79e+03	&	1.38e+03	&	2.06e+01	&	5764	&	--- 	&	2	&	1.7041	&	128.78	&	-13.75	&	n	&	eb\\
80025900	&	4.0594	&	6.71e+03	&	6.28e+02	&	6.11e+01	&	3876	&	---	    &	2	&	8.6198	&	121.00	&	-46.33	&	y	&	eb\\
231972938	&	1.5507	&	6.69e+03	&	5.75e+02	&	2.75e+01	&	6210	&	4.35	&	1	&	3.9200	&	95.47	&	-46.21	&	y	&	eb\\
142677613	&	3.0912	&	6.64e+03	&	9.24e+02	&	5.28e+01	&	8357	&	4.16	&	2	&	16.442	&	103.60	&	-14.45	&	y	&	eb\\
98064760	&	9.9000	&	6.60e+03	&	4.87e+02	&	7.64e+01	&	---	    &	--- 	&	1	&	6.7401	&	289.91	&	-40.81	&	y	&	eb\\
292293385	&	0.2694	&	6.50e+03	&	3.42e+04	&	5.26e+00	&	5492	&	4.48	&	1	&	3.3371	&	278.49	&	-69.86	&	n	&	usp\\
167795859	&	3.1061	&	6.50e+03	&	2.75e+02	&	4.98e+01	&	5479	&	4.00	&	1	&	16.795	&	104.42	&	-61.75	&	y	&	eb\\
387463207	&	0.1899	&	6.48e+03	&	9.64e+06	&	5.63e+00	&	6259	&	4.21	&	1	&	27.263	&	42.46	&	8.93	&	n	&	usp\\
437230910	&	0.3805	&	6.39e+03	&	3.01e+03	&	1.48e+01	&	5406	&	4.10	&	1	&	0.2367	&	164.77	&	-21.70	&	y	&	pl\\
396934949	&	0.1887	&	6.36e+03	&	1.74e+04	&	6.15e+00	&	4340	&	4.29	&	1	&	4.5204	&	63.47	&	3.31	&	n	&	usp\\
96106862	&	5.3800	&	6.32e+03	&	6.05e+02	&	5.80e+01	&	6554	&	4.03	&	2	&	1.2424	&	147.20	&	-8.49	&	n	&	eb\\
146601318	&	0.4935	&	6.27e+03	&	1.21e+03	&	1.39e+01	&	5257	&	3.75	&	1	&	1.9066	&	149.25	&	-1.35	&	y	&	pl\\
456609929	&	4.5678	&	6.23e+03	&	1.15e+04	&	1.59e+01	&	7643	&	4.09	&	2	&	5.5403	&	282.35	&	-59.42	&	y	&	eb\\
130942594	&	10.792	&	6.23e+03	&	1.09e+03	&	1.32e+02	&	6362	&	--- 	&	1	&	20.078	&	188.34	&	-32.74	&	y	&	eb\\
87260967	&	0.5452	&	6.22e+03	&	7.45e+03	&	1.40e+01	&	4665	&	---	    &	1	&	0.1441	&	272.76	&	-47.51	&	y	&	pl\\
303811839	&	6.9220	&	6.18e+03	&	4.12e+02	&	7.14e+01	&	5938	&	--- 	&	1	&	2.0336	&	164.54	&	-55.98	&	n	&	eb\\
220402290	&	0.7852	&	6.12e+03	&	2.36e+03	&	1.53e+01	&	5817	&	4.36	&	2	&	4.5066	&	69.57	&	-57.20	&	n	&	eb\\
271225404	&	2.1195	&	6.11e+03	&	5.67e+02	&	3.20e+01	&	3764	&	4.62	&	2	&	47.344	&	211.10	&	-31.31	&	y	&	eb\\
381948745	&	1.5953	&	6.02e+03	&	4.78e+03	&	1.37e+01	&	7084	&	3.55	&	1	&	7.1065	&	76.36	&	-56.30	&	n	&	eb\\
128679896	&	11.130	&	5.99e+03	&	8.56e+04	&	2.35e+02	&	5594	&	4.20	&	2	&	19.882	&	299.25	&	-43.61	&	y	&	eb\\
394657952	&	0.8723	&	5.94e+03	&	1.77e+03	&	2.02e+01	&	5725	&	---	    &	1	&	2.7255	&	158.68	&	-77.96	&	y	&	eb\\
105706808	&	0.1986	&	5.93e+03	&	8.48e+03	&	7.58e+00	&	---	    &	--- 	&	1	&	0.4872	&	156.25	&	-39.51	&	n	&	usp\\
231840927	&	1.6357	&	5.87e+03	&	1.46e+03	&	2.17e+01	&	5726	&	4.44	&	2	&	0.4241	&	25.17	&	-65.41	&	y	&	eb\\
231833061	&	23.335	&	5.85e+03	&	1.17e+04	&	2.31e+02	&	4580	&	4.34	&	2	&	48.949	&	24.00	&	-65.69	&	y	&	eb\\
294328887	&	0.3545	&	5.79e+03	&	1.89e+03	&	8.30e+00	&	3079	&	4.85	&	1	&	10.246	&	108.97	&	-59.34	&	n	&	eb\\
124712820	&	0.8256	&	5.72e+03	&	1.08e+03	&	1.69e+01	&	3989	&	4.34	&	2	&	1.2622	&	104.76	&	-11.72	&	n	&	eb\\
312907176	&	0.7580	&	5.71e+03	&	1.21e+03	&	1.47e+01	&	3260	&	5.00	&	1	&	43.759	&	182.91	&	-68.82	&	y	&	eb\\
307033218	&	25.755	&	5.66e+03	&	1.92e+02	&	7.51e+01	&	5216	&	4.08	&	1	&	42.014	&	123.30	&	-69.32	&	y	&	eb\\
438682877	&	0.9221	&	5.41e+03	&	1.65e+03	&	1.56e+01	&	5820	&	--- 	&	2	&	0.7677	&	202.22	&	-44.68	&	n	&	eb\\
59038880	&	0.3053	&	5.29e+03	&	8.22e+07	&	4.70e+00	&	5325	&	4.62	&	2	&	79.385	&	57.36	&	12.90	&	n	&	eb\\
412942849	&	0.3806	&	5.18e+03	&	6.39e+03	&	8.75e+00	&	6222	&	3.43	&	1	&	0.5721	&	204.57	&	-55.57	&	n	&	pl\\
168697816	&	0.6920	&	5.17e+03	&	1.91e+03	&	1.13e+01	&	6456	&	4.37	&	2	&	2.4111	&	61.66	&	-34.96	&	n	&	eb\\
266816672	&	0.3031	&	5.03e+03	&	7.30e+06	&	4.92e+00	&	6064	&	4.32	&	2	&	21.483	&	30.93	&	-19.88	&	n	&	eb\\
75818403	&	0.5635	&	4.94e+03	&	7.18e+03	&	6.93e+00	&	3669	&	4.75	&	2	&	62.353	&	139.52	&	-46.08	&	n	&	eb\\
302038263	&	1.5474	&	4.93e+03	&	8.26e+02	&	1.96e+01	&	--- 	&	---	    &	2	&	6.0819	&	131.95	&	-69.39	&	n	&	eb\\
23328912	&	2.5639	&	4.83e+03	&	7.65e+02	&	3.80e+01	&	8690	&	4.15	&	2	&	0.8293	&	167.44	&	-35.40	&	n	&	eb\\
1190662 	&	0.4440	&	4.79e+03	&	1.30e+03	&	1.76e+01	&	3595	&	4.28	&	2	&	2.3930	&	136.03	&	-16.49	&	y	&	eb\\
125501325	&	4.3858	&	4.77e+03	&	7.40e+02	&	4.11e+01	&	--- 	&	---	    &	2	&	5.9093	&	105.85	&	-10.70	&	n	&	eb\\
13738984	&	0.2769	&	4.76e+03	&	1.69e+04	&	6.68e+00	&	5818	&	4.35	&	2	&	0.5887	&	133.70	&	-11.03	&	n	&	usp\\
89787749	&	0.5413	&	4.75e+03	&	6.98e+02	&	1.68e+01	&	3374	&	4.91	&	1	&	23.436	&	127.87	&	-59.20	&	n	&	eb\\
92938279	&	0.2750	&	4.66e+03	&	2.92e+05	&	4.38e+00	&	3327	&	4.89	&	2	&	18.534	&	83.63	&	-22.64	&	n	&	usp\\
92938280	&	0.2750	&	4.54e+03	&	1.20e+05	&	4.45e+00	&	3396	&	4.95	&	2	&	15.406	&	83.63	&	-22.64	&	n	&	usp\\
29847695	&	4.9980	&	4.46e+03	&	7.29e+02	&	5.01e+01	&	6375	&	--- 	&	2	&	7.6000	&	23.58	&	-8.45	&	y	&	eb\\
350740905	&	0.1555	&	4.37e+03	&	1.88e+05	&	4.22e+00	&	8181	&	3.86	&	1	&	0.8493	&	88.55	&	-56.17	&	n	&	usp\\
436243444	&	0.6793	&	4.33e+03	&	9.70e+02	&	1.70e+01	&	5093	&	4.24	&	1	&	3.2989	&	84.13	&	8.43	&	y	&	eb\\
396953669	&	0.3327	&	4.30e+03	&	7.56e+03	&	6.55e+00	&	4719	&	---	    &	2	&	0.3230	&	63.88	&	3.36	&	n	&	eb\\
30287190	&	3.1243	&	4.21e+03	&	3.17e+02	&	4.35e+01	&	6217	&	4.16	&	2	&	2.6468	&	207.60	&	-31.23	&	n	&	eb\\
2353789 	&	1.8718	&	4.20e+03	&	8.55e+02	&	1.98e+01	&	6969	&	3.91	&	1	&	3.0641	&	134.27	&	-28.77	&	n	&	eb\\
279454212	&	11.762	&	4.17e+03	&	2.06e+02	&	7.62e+01	&	5951	&	4.42	&	1	&	4.1042	&	58.47	&	-16.95	&	y	&	eb\\
425206178	&	0.1647	&	4.07e+03	&	1.41e+04	&	4.97e+00	&	--- 	&	--- 	&	1	&	6.2160	&	111.55	&	7.68	&	n	&	usp\\
161570793	&	0.2762	&	4.04e+03	&	4.21e+05	&	4.29e+00	&	6262	&	4.37	&	2	&	4.9210	&	74.44	&	-42.89	&	n	&	usp\\
269762258	&	1.3247	&	3.99e+03	&	4.10e+04	&	2.63e+01	&	3358	&	4.67	&	2	&	11.874	&	66.47	&	-81.31	&	y	&	eb\\
20095466	&	0.2118	&	3.97e+03	&	2.75e+03	&	7.97e+00	&	5966	&	4.49	&	1	&	0.2755	&	89.22	&	-36.07	&	y	&	usp\\
284729177	&	0.2068	&	3.88e+03	&	6.91e+04	&	4.44e+00	&	7571	&	3.43	&	1	&	1.4845	&	67.66	&	10.60	&	n	&	usp\\
185805011	&	2.0054	&	3.78e+03	&	7.87e+02	&	1.85e+01	&	3477	&	4.83	&	2	&	27.171	&	130.69	&	-31.03	&	y	&	eb\\
214299966	&	3.0675	&	3.64e+03	&	5.70e+02	&	3.30e+01	&	6181	&	4.14	&	1	&	1.7140	&	339.16	&	-36.94	&	n	&	pl\\
147102369	&	2.0638	&	3.46e+03	&	4.47e+02	&	3.45e+01	&	7231	&	3.84	&	2	&	5.8315	&	124.80	&	-30.42	&	y	&	eb\\
381854774	&	1.6447	&	3.45e+03	&	7.77e+02	&	2.21e+01	&	6375	&	4.12	&	1	&	1.0266	&	321.79	&	-54.34	&	y	&	pl\\
178889597	&	1.8842	&	3.43e+03	&	7.90e+02	&	2.99e+01	&	4777	&	---	    &	1	&	2.4647	&	195.12	&	-38.84	&	y	&	eb\\
282051803	&	2.4625	&	3.41e+03	&	1.00e+03	&	2.24e+01	&	6246	&	3.87	&	2	&	7.0610	&	132.22	&	-75.92	&	y	&	eb\\
186544053	&	8.9117	&	3.41e+03	&	2.15e+02	&	4.64e+01	&	---	    &	4.67	&	1	&	1.7649	&	132.92	&	-33.10	&	y	&	pl\\
198008281	&	0.2631	&	3.36e+03	&	3.04e+06	&	4.21e+00	&	7089	&	4.04	&	2	&	3.2852	&	60.26	&	-54.40	&	n	&	usp\\
33689349	&	3.4974	&	3.25e+03	&	6.85e+02	&	2.50e+01	&	7098	&	4.10	&	2	&	24.721	&	92.93	&	-17.68	&	y	&	eb\\
406276109	&	7.5474	&	3.23e+03	&	3.26e+02	&	4.51e+01	&	3200	&	4.82	&	2	&	10.415	&	325.64	&	-64.21	&	n	&	eb\\
243391171	&	1.5167	&	3.18e+03	&	4.65e+03	&	1.52e+01	&	5943	&	---	    &	1	&	2.3712	&	205.06	&	-45.70	&	n	&	eb\\
335452175	&	7.7501	&	3.16e+03	&	4.01e+03	&	6.51e+01	&	--- 	&	---	    &	1	&	0.8537	&	226.27	&	-47.05	&	n	&	pl\\
360571125	&	2.4626	&	3.13e+03	&	7.99e+02	&	2.53e+01	&	9517	&	3.60	&	2	&	1.1710	&	263.87	&	-55.76	&	y	&	eb\\
167007869	&	0.5509	&	3.10e+03	&	2.12e+03	&	1.06e+01	&	4949	&	--- 	&	1	&	0.1814	&	94.80	&	-67.59	&	n	&	misc\\
9433212 	&	5.0138	&	3.07e+03	&	1.45e+02	&	3.70e+01	&	5878	&	4.17	&	2	&	4.2410	&	73.54	&	-9.49	&	n	&	eb\\
403136932	&	0.3028	&	3.05e+03	&	4.15e+05	&	3.93e+00	&	3391	&	4.82	&	2	&	46.579	&	333.76	&	-54.27	&	n	&	eb\\
332003916	&	0.6461	&	3.02e+03	&	1.71e+04	&	5.50e+00	&	7458	&	3.54	&	4	&	0.7006	&	179.10	&	-48.79	&	n	&	misc\\
349523518	&	0.1977	&	3.02e+03	&	2.61e+04	&	3.93e+00	&	6172	&	4.07	&	1	&	3.6427	&	111.49	&	-63.22	&	n	&	usp\\
277912443	&	0.5379	&	3.01e+03	&	2.10e+03	&	9.50e+00	&	5964	&	4.04	&	1	&	0.6526	&	354.44	&	-73.91	&	n	&	misc\\
129781743	&	0.2012	&	3.01e+03	&	9.79e+03	&	5.36e+00	&	6655	&	4.24	&	1	&	0.7832	&	36.24	&	-38.68	&	n	&	usp\\
350518984	&	10.367	&	3.00e+03	&	1.40e+02	&	6.01e+01	&	7029	&	---	    &	1	&	2.8731	&	86.18	&	-56.71	&	y	&	eb\\
72229825	&	1.7954	&	2.99e+03	&	1.78e+02	&	2.11e+01	&	6415	&	4.41	&	1	&	1.5651	&	186.75	&	-43.67	&	n	&	pl\\
350479101	&	5.9589	&	2.98e+03	&	2.65e+02	&	5.77e+01	&	6662	&	4.05	&	1	&	2.7658	&	85.79	&	-59.14	&	y	&	eb\\
272086869	&	3.9823	&	2.97e+03	&	3.02e+02	&	3.20e+01	&	8799	&	4.03	&	2	&	1.7775	&	115.67	&	-72.53	&	y	&	eb\\
349154435	&	4.4313	&	2.96e+03	&	2.13e+02	&	3.36e+01	&	5832	&	4.19	&	2	&	4.1264	&	108.18	&	-62.01	&	y	&	eb\\
349911034	&	0.5385	&	2.92e+03	&	9.12e+02	&	1.17e+01	&	5799	&	4.20	&	1	&	1.6651	&	113.97	&	-65.68	&	y	&	pl\\
270427198	&	2.5609	&	2.85e+03	&	2.07e+04	&	4.40e+00	&	6984	&	3.35	&	3	&	0.7279	&	96.35	&	-78.57	&	n	&	misc\\
255567460	&	6.8989	&	2.84e+03	&	2.13e+02	&	4.38e+01	&	9984	&	4.35	&	1	&	1.3900	&	96.44	&	-53.10	&	y	&	pl\\
312068854	&	7.1007	&	2.82e+03	&	1.38e+02	&	5.83e+01	&	5716	&	4.31	&	1	&	1.5401	&	140.02	&	8.44	&	y	&	pl\\
52382500	&	0.1968	&	2.80e+03	&	5.32e+04	&	3.97e+00	&	5944	&	3.61	&	1	&	1.7025	&	24.19	&	-69.95	&	n	&	usp\\
48318875	&	0.3028	&	2.76e+03	&	7.34e+04	&	4.24e+00	&	5873	&	3.99	&	2	&	4.2039	&	223.30	&	-31.62	&	n	&	eb\\
24347173	&	0.3516	&	2.75e+03	&	1.57e+03	&	1.16e+01	&	5744	&	4.22	&	2	&	0.8472	&	79.17	&	-8.70	&	n	&	eb\\
289596375	&	7.2587	&	2.72e+03	&	4.74e+02	&	4.22e+01	&	6170	&	3.59	&	2	&	2.7868	&	131.79	&	-13.06	&	y	&	eb\\
92743594	&	8.4461	&	2.70e+03	&	2.92e+02	&	3.26e+01	&	4835	&	--- 	&	4	&	1.9285	&	129.00	&	-53.61	&	y	&	eb\\
453828066	&	9.1358	&	2.67e+03	&	4.58e+02	&	4.04e+01	&	6545	&	4.08	&	1	&	9.3237	&	53.36	&	0.11	&	n	&	eb\\
166319733	&	0.1410	&	2.66e+03	&	5.82e+04	&	4.09e+00	&	5425	&	3.86	&	1	&	5.0870	&	206.68	&	-37.76	&	n	&	usp\\
372909935	&	3.6118	&	2.64e+03	&	2.09e+02	&	3.64e+01	&	7015	&	4.12	&	1	&	1.1117	&	119.45	&	-64.21	&	y	&	pl\\
244252435	&	0.9435	&	2.62e+03	&	2.45e+04	&	3.39e+01	&	3675	&	4.57	&	2	&	5.4926	&	74.37	&	-3.95	&	n	&	eb\\
23555025	&	4.7455	&	2.62e+03	&	2.34e+02	&	3.59e+01	&	6001	&	4.34	&	1	&	4.1329	&	169.03	&	-32.65	&	n	&	eb\\
362086333	&	0.2828	&	2.61e+03	&	9.33e+03	&	5.41e+00	&	10440	&	4.18	&	1	&	1.0040	&	107.90	&	-84.46	&	n	&	usp\\

\hline\hline 
\end{longtable}

\bsp
\label{lastpage}

\begin{thebibliography}{99}
	
\bibitem[\protect\citeauthoryear{Ballard}{2019}]{ballard:2019} 
Ballard, S., 2019, AJ, 157, 113
\bibitem[\protect\citeauthoryear{Ballesteros et al.}{2018}]{ballesteros:2018} 
Ballesteros, F.~J., Arnalte-Mur, P., Fernandez-Soto, A., Mart\'inez, V.~J., 2018, MNRAS, 473, 21
\bibitem[\protect\citeauthoryear{Barclay et al.}{2018}]{barclay:2018} 
Barclay, T., Pepper, J. \& Quintana, E., 2018, ApJS, 239, 2
\bibitem[\protect\citeauthoryear{B\'eky et al.}{2014}]{beky:2014} 
B\'eky, B., Holman, M.~J., Kipping, D.~M., Noyes, R.~W., 2014, ApJ, 788, 1
\bibitem[\protect\citeauthoryear{Bodman \& Quillen}{2016}]{bodman:2016} 
Bodman, E.~H.~L. \& Quillen, A., 2016, ApJ, 819, 34
\bibitem[\protect\citeauthoryear{Bouma et al.}{2017}]{bouma:2017} 
Bouma, L.~G., Winn, J.~N., Kosiarek, J., McCullough, P.~R., 2017, arXiv e-prints:1705.08891
\bibitem[\protect\citeauthoryear{Boyajian et al.}{2016}]{boyajian:2016} 
Boyajian, T.~S., LaCourse, D.~M., Rappaport, S.~A. et al., 2016, MNRAS, 457, 3988
\bibitem[\protect\citeauthoryear{Foreman-Mackey et al.}{2016}]{dfm:2016} 
Foreman-Mackey, D., Morton, T.~D., Hogg, D.~W., Agol, E., Sch\"olkopf, B., 2016, AJ, 152, 206
\bibitem[\protect\citeauthoryear{Foukal}{2017}]{foukal:2017} 
Foukal, P., 2017, ApJ, 842, 3
\bibitem[\protect\citeauthoryear{Giles \& Walkowicz}{2019}]{giles:2019} 
Giles, D. \& Walkowicz, L., 2019, MNRAS, 484, 834
\bibitem[\protect\citeauthoryear{Jansen \& Kipping}{2018}]{jansen:2018} 
Jansen, T. \& Kipping, D., 2018, MNRAS, 478, 3025
\bibitem[\protect\citeauthoryear{Jayasinghe et al.}{2018}]{jayasinghe:2018}
Jayasinghe, T., Kochanek, C.~S., Stanek, K.~Z., Shappee, B.~J., Holoien, T.W., Thompson, T.~A., Prieto, J.~L., Dong, S., Pawlak, M., Shields, J.~V., Pojmanski, G., Otero, S., Britt, C.~A., Will, D., 2018, MNRAS, 477, 3145
\bibitem[\protect\citeauthoryear{Jayasinghe et al.}{2019}]{jayasinghe:2019}
Jayasinghe, T., Stanek, K.~Z., Kochanek, C.~S., Shappee, B.~J., Holoien, T.W., Thompson, T.A., Prieto, J.L., Dong, S., Pawlak, M., Peicha, O., Shields, J.~V., Pojmanski, G., Otero, S., Britt, C.~A., Will, D., 2019, MNRAS, 486, 1907
\bibitem[\protect\citeauthoryear{Katz}{2017}]{katz:2017} 
Katz, J.~I., 2017, MNRAS, 471, 3680
\bibitem[\protect\citeauthoryear{Kawahara \& Masuda}{2019}]{kawahara:2019} 
Kawahara, H. \& Masuda, K., 2019, ApJ, 157, 218
\bibitem[\protect\citeauthoryear{Kipping et al.}{2013}]{kipping:2013} 
Kipping, D.~M., Hartman, J., Buchhave, L.~A., Schmitt, A.~R., Bakos, G. \'A.
\& Nesvorn\'y, D., 2013, ApJ, 770, 101
\bibitem[\protect\citeauthoryear{Kov\'acs et al.}{2002}]{kovacs:2002} 
Kov\'acs, G., Zucker, S., \& Mazeh, T., 2002, A\&A, 391, 369
\bibitem[\protect\citeauthoryear{Kov\'acs et al.}{2006}]{kovacs:2006} 
Kov\'cs, G., Zucker, S., \& Mazeh, T., 2016, BLS: Box-fitting Least Squares,
Astrophysics Source Code Library
\bibitem[\protect\citeauthoryear{Martinez et al.}{2019}]{martinez:2019} 
Martinez, M.~A.~S., Stone, N.~C., Metzger, B.~D., 2019, MNRAS, 489, 5119
\bibitem[\protect\citeauthoryear{Neslusan \& Budaj}{2017}]{nelusan:2017} 
Neslusan, L. \& Budaj, J. 2017, A\&A, 600, 86
\bibitem[\protect\citeauthoryear{Parks et al.}{2014}]{parks:2014}
Parks, J.~R., Plavchan, P., White, R.~J., Gee, A.~H., 2014, ApJS, 211, 1
\bibitem[\protect\citeauthoryear{Plavchan et al.}{2008}]{plavchan:2008}
Plavchan, P., Jura, M., Kirkpatrick, J.~D., Cutri, R.~M., Gallagher, S.~C., 2008, ApJS, 175, 1
\bibitem[\protect\citeauthoryear{Prsa}{2020}]{prsa:2020} 
Prsa, A., 2020, ``Eclipsing binary stars: Indispensableastrophysical labs'',
TESS Program G011154, last accessed 24 June 2020
https://heasarc.gsfc.nasa.gov/docs/tess/data/approved-programs/G011154.txt 
\bibitem[\protect\citeauthoryear{Ricker et al.}{2015}]{ricker:2015} 
Ricker G.~R. et al., 2015, J. Astron. Telesc. Instrum. Syst., 1, 014003
\bibitem[\protect\citeauthoryear{Schmidt}{2019}]{schmidt:2019} 
Schmidt, E.~G., 2019, ApJ, 880, 7
\bibitem[\protect\citeauthoryear{Schmitt}{2017}]{schmitt:2017} 
Schmitt, J.~R., Jenkins, J.~M. \& Fischer, D.~A., 2017, ApJ, 153, 180
\bibitem[\protect\citeauthoryear{Shappee}{2014}]{shappee:2014}
Shappee, B.~J., Prieto, J.~L., Grupe, D., Kochanek, C.~S., Stanek, K.~Z., De Rosa, G., Mathur, S., Zu, Y., Peterson, B.~M., Pogge, R.~W., Komossa, S., Im, M., Jencson, J., Holoien, T.~W., Basu, U., Beacorn, J.~F, Szczygiel, D.~M., Brimacombe, J., Adams, S., Campillay, A., Choi, C., Contreras, C., Dietrich, M., Dubberley, M., Elphick, M., Foale, S., Giustini, M., Gonzalez, C., Hawkins, E., Howell, D.~A., Hsiao, E.~Y., Koss, M., Leighly, K.~M., Morrell, N., Mudd, D., Mullins, D., Nugent, J.~M., Parrent, J., Phillips, M.~M., Pojmanski, G., Rosing, W., Ross, R., Sand, D., Terndrup, D.~M., Valenti, S., Walker, Z., Yoon, Y., 2014, ApJ, 788, 1
\bibitem[\protect\citeauthoryear{Stassun et al.}{2018}]{stassun:2018} 
Stassun, K.~G., Oelkers, R.~J., Pepper, J., et al., 2018, AJ, 156, 102 
\bibitem[\protect\citeauthoryear{Sucerquia et al.}{2019}]{sucerquia:2019} 
Sucerquia, M., Alvarado-Montes, J.~A., Zuluaga, J.~I., Cuello, N., Giuppone, C.,
2019, MNRAS, 489, 2313
\bibitem[\protect\citeauthoryear{Sullivan et al.}{2015}]{sullivan:2015} 
Sullivan, P.~W., Winn, J.~N., Berta-Thompson, Z.~K. et al., 2015, ApJ, 809, 77
\bibitem[\protect\citeauthoryear{Sullivan et al.}{2017}]{sullivan:2017} 
Sullivan, P.~W., Winn, J.~N., Berta-Thompson, Z.~K. et al., 2017, ApJ, 837, 99
\bibitem[\protect\citeauthoryear{Uehara et al.}{2016}]{uehara:2016} 
Uehara, S., Kawahara, H., Masuda, K., Yamada, S. \& Aizawa, M., 2016, ApJ, 822, 2
\bibitem[\protect\citeauthoryear{Wang et al.}{2015}]{wang:2015} 
Wang, J., Fischer, D.~A., Barclay, T., et al., 2015, ApJ, 815, 127
\bibitem[\protect\citeauthoryear{Watson et al.}{2006}]{watson:2006}
Watson, C.~L., Henden, A.~A., \& Price, A., 2006, The Society for Astronomical Sciences 25th Annual Symposium on Telescope Science, p.47
\bibitem[\protect\citeauthoryear{Wheeler \& Kipping}{2019}]{wheeler:2019} 
Wheeler, A. \& Kipping, D., 2019, MNRAS, 485, 5498
\bibitem[\protect\citeauthoryear{Wright et al.}{2016}]{wright:2016} 
Wright, J.~T. \& Sigurdsson, S., 2016, ApJL, 829, L3
\bibitem[\protect\citeauthoryear{Wyatt et al.}{2018}]{wyatt:2018} 
Wyatt, M.~C., van Lieshout, R., Kennedy, G.~M. \& Boyajian, T.~S., 2018, MNRAS, 473, 5286
\end{thebibliography}
\end{document}